\documentclass[preprint2]{aastex}
\usepackage[T1]{fontenc}
\usepackage[latin9]{inputenc}
\usepackage{amsmath}
\usepackage{graphicx}
\usepackage{amssymb}
\usepackage{verbatim}
\usepackage{fixltx2e}
\begin{document}

\title{Dithering Strategies and Point-Source Photometry}

\author{Johan Samsing}
\affil{Dark Cosmology Centre, Niels Bohr Institute, Copenhagen, Denmark}
\author{Alex G.\ Kim}
\affil{Physics Division, Lawrence Berkeley National Laboratory, Berkeley CA 94720}

\begin{abstract}
The accuracy in the photometry of a point source depends on the point-spread function (PSF), detector pixelization, and observing strategy.  The PSF and pixel response describe the spatial blurring of the source, the pixel scale describes the spatial sampling of a single exposure, and the observing strategy determines the set of dithered exposures with pointing offsets from which the source flux is inferred.  In a wide-field imaging survey, sources of interest are randomly distributed within the field of view and hence are centered randomly within a pixel. A given hardware configuration and observing strategy therefore have a distribution of photometric uncertainty for sources of fixed flux
that fall in the field.
In this article we explore the ensemble behavior of photometric and position accuracies for different PSFs, pixel scales, and dithering patterns.
We find that the average uncertainty in the
flux determination depends slightly on dither strategy, whereas
the position determination can be strongly dependent on the dithering.  
For cases with pixels much larger than the PSF, the uncertainty
distributions can be non-Gaussian, with rms values that are particularly sensitive to the dither strategy.
We also find that for these configurations with large pixels, pointings dithered by a fractional pixel amount
do not always give minimal average uncertainties; this is in contrast to image reconstruction for which fractional dithers are optimal.
When fractional pixel dithering is favored, a pointing accuracy of better than $\sim 0.15$ pixel width is required to maintain half the 
advantage over random dithers.
\ \\
\ \\
\end{abstract}

\section{INTRODUCTION}
Survey imagers are designed to provide accurate measurements of multiple objects in a single exposure.  Given a fixed number of detector
pixels, the choice of pixel scale determines the field of view and influences the angular resolution.  Optimization of the pixel scale trades the
multiplex advantage of simultaneous observation of many sources with the accuracy of source flux, position, and shape measurements.

With a space telescope the point-spread function (PSF) can be designed to be stable over multiple exposures, then
the observing strategy can be used to
affect the measurement accuracy.  Dithering breaks an observation into a sequence of exposures with subpixel pointing offsets
to recover Nyquist sampling from images that are individually undersampled.  The dithering approach allows having the large field of view from
angularly coarse pixels while still allowing robust point-source
photometry \citep{1999PASP..111.1434L}  and the measurement of subpixel spatial structure
of extended objects such as galaxies \citep{1999PASP..111..227L,2002PASP..114..144F,2002PASP..114...98B}. Subdivision of a single pointing into multiple exposures
can be done efficiently as long as the short exposures are not dominated by detector read noise, have relatively long exposure times compared with readout time, and do not
produce data volumes that exceed data storage and telemetry constraints.

The WFIRST mission set forward in the Astro2010 report is a satellite experiment that requires accurate measurements of many objects
within its field of view.  WFIRST measures shapes and colors of many galaxies to detect shear caused by  mass inhomogeneities using $\sim 50$
galaxies per square arcminute.
WFIRST measures the time-evolving brightness of stars in dense fields toward the Galactic bulge to search for planets, and
thousands of Type Ia supernovae to map the expansion history of the universe.  The fundamental measurement in these latter
two cases is the flux of a point source.

In this article we explore how point-source photometry for an ensemble of sources drives the design and observing strategy of a space-based mission.
The ensemble behavior is of interest
because sources are randomly distributed in the sky and are therefore randomly positioned within a pixel;
this is particularly relevant when the pixel is much larger than the PSF.
The design parameter of interest is the pixel scale, which can be chosen so that a pixel is smaller, similar, or much larger in size
compared with the PSF.  The photometric accuracy depends on the number of dither steps and the dither-pointing grid.  Although Nyquist
sampling with a uniform dither pattern is optimal for image reconstruction \citep{1999PASP..111..227L}, this is not necessarily true for
point-source photometry.  In situations where a uniform dither pattern is advantageous, we determine the pointing accuracy required to maintain
that advantage.  We consider both cases where the point-source position is independently known or must be derived from the data. 

Our study applies to the photometry of a single pointing of a microlensed star or supernova.
In WFIRST, the star and supernova fields are observed hundreds of times with
random subpixel pointings.
For supernovae,
the underlying host-galaxy structure
can be measured to the Nyquist frequency
using data from all visits.  Our treatment operates as if the host-galaxy surface brightness
is determined independently and is subtracted from the images of the visit of interest.
Similarly, the centroid position of stars and supernovae can be determined from the multiple visits that constitute the light curves.
The position can be considered to be independently known in the analysis of any particular image.  A rigorous treatment simultaneously
fits for the position, background, and fluxes of all observations.

Our analysis is based on the Fisher matrix approach, which is a way to analytically estimate parameter uncertainties and correlations to first order
without mapping the likelihood surface for each fit. 

The paper is organized as follows.  In \S\ref{PSF:sec}, we review PSF photometry and the calculation of
its uncertainty.  Calculations of the mean and variance of the photometric uncertainty
for different choices of native PSF, pixel scale, and number of dithered exposures are given in \S\ref{uncertainties:sec}.
The effect of cosmic rays is shown in \S\ref{cosmic:sec}.
We summarize with conclusions in \S\ref{conclusions:sec}.

\section{POINT-SOURCE PHOTOMETRY AND UNCERTAINTIES}
\label{PSF:sec}
This section presents our model for the PSF and pixelized data, and demonstrates how PSF photometry is performed to estimate flux and position uncertainties from a fit to the data.

\subsection{Point-Spread Function and Effective Point-Spread Function} 
\label{ePSF:sec}
Extragalactic supernovae are sufficiently small and
distant to be considered pointlike (i.e., a $\delta$ function)
when its light reaches Earth.  The shape of the supernova signal within the detector
(the PSF) is the convolution of the blur contributions from atmospheric scattering,
spacecraft jitter, telescope diffraction and wavefront error, and  detector diffusion.  Furthermore,
the detector is pixelized so the measured signals are a discrete sampling
of the convolution of the PSF and pixel-response function, the effective PSF (ePSF or $P$).    For the space-based mission considered in this article, the main
contributions to the PSF are the diffraction due to the telescope
and the charge diffusion within the detector. 

The diffraction for a telescope with an unobscured circular aperture is an Airy disk described by the intensity pattern
\begin{equation}
I(q)=I_{0}\left(\frac{2J_{1}(\pi q)}{\pi q}\right)^{2},
\end{equation}
where $J_{1}$ is the Bessel function of the first kind of order one and
$q$ the distance in units of $(l_{foc}\lambda/D)$ on the chip from
the centroid. Here $l_{foc}$ is the telescope focal length, $D$
the diameter of the telescope mirror and $\lambda$ the wavelength
of observation.
Detector diffusion of
the fully-depleted detectors under consideration is described by a
Gaussian profile $\mathcal{N}$ with width, $\sigma_{diff}$.   (In our formalism, the charge diffusion term can represent all sources of Gaussian blur.)
The ePSF accounts for the pixelization by convolving the PSF with the pixel-response function as predetermined by calibration.  In our calculations we take the pixel response to be a two-dimensional boxcar function $\Pi$, although, in general, the response could have intrapixel variation.  The ePSF is  $P=I \otimes \mathcal{N} \otimes \Pi$.

We assume that the ePSF is  derived from field objects and extensive
pre- and in-flight calibration and contributes negligible statistical uncertainty to the
PSF photometry.  This translates into an experimental requirement for
the amount of calibration data needed to model the ePSF: a  challenging but
feasible task considering the stability of the space platform and the
surface density of point sources on each image.  The ePSF calibration includes
contributions from possible intrapixel variation.

\subsection{The Data}\label{sec:thedata}
The data for
a single visit are the counts from each pixel of each of the exposures
in the dither sequence. For a $d\times d$ dither grid there are a total
of $d^{2}$ exposures. To directly compare the relative performance of
different dither sequences, the total exposure time for each visit is fixed to $t_{tot}$ so that each individual dither position gets an exposure
time $t_{tot}/d^{2}$. 

The data noise is taken to have contributions from the sky background, the source, and readout noise.  Dark current is not explicitly considered:
for a given pixel scale it can be included with the sky background.
Denoting the sky counts
per unit steradian $n_{s}$, the side length of a square
pixel $a$, the flux $f$, 
and the readout noise per pixel $R$, the variance of the data from pixel $\beta$ in exposure $\alpha$
takes the form
\begin{equation}
\sigma_{\alpha\beta}^{2}=\frac{t_{tot}}{d^{2}}\left[n_{s}\left(\frac{a}{l_{foc}}\right)^{2}+fP{}_{\alpha\beta}(x,y)\right]+R^{2},
\label{eq:20}
\end{equation}
where $P$ is the ePSF defined in \S\ref{ePSF:sec}.
We use the small-angle approximation since the pixel sizes are much smaller than the focal length. 
The noise between different
pixels is taken to be uncorrelated. 

\subsection{PSF Photometry and Uncertainties}
Point-source PSF photometry fits data from pixels, each centered at position $x$ and $y$, to a model $f P(x-x_0,y-y_0)$.
The fit parameters
$\mathbf{p}=\{f,x_{0},y_{0}\}$ include $f$ for the flux in counts s$^{-1}$
and $x_{0}$ and $y_{0}$ for the centroid position. We consider cases where the centroid
position is independently known or must also be derived from the fit to the data.
The flux and ePSF are assumed to not evolve in the short interval
that the dither sequence is performed, although 
 it can generally vary from exposure to exposure.

The most probable values for the vector $\mathbf{p}$ are found by maximizing the likelihood or minimizing $\chi^{2}$
when the noise is Gaussian. Working from the Fisher information matrix (FIM)
formalism it can be shown that $\sigma_{i}^{2}\ge\left(F^{-1}\right)_{ii}$;
a calculation of the Fisher elements may give precise information
of how well parameter $i$ is estimated.
In general the FIM takes the form
\begin{eqnarray}
F_{ij}=\frac{\partial O^{T}}{\partial p_{i}}C^{-1}\frac{\partial O}{\partial p_{j}}+\frac{1}{2}Tr\left(C^{-1}\frac{\partial C}{\partial p_{i}}C^{-1}\frac{\partial C}{\partial p_{j}}\right),
\label{eq:200}
\end{eqnarray}
where $O$ is a vector with the observables, $C$ is the corresponding covariance matrix, and $p_{i}$ is parameter $i$.
When the noise is not dependent on the parameters we fit for, the second term is simply zero. This is the case
for the sky-noise limit, but not in the source-noise limit.  We consider only the first term in equation \ref{eq:200}  in our analysis, which is appropriate in the source-noise limit when the number of measured photons is significantly greater than the number of used pixels. As written in \S \ref{sec:thedata}, we assume no correlation between our observables: i.e., we take $C$ to be diagonal.
Keeping only the first term and inserting the observables $(fP_{\alpha \beta})$ with the corresponding errors $(\sigma_{\alpha \beta})$ the FIM in our case for a single epoch takes the following form:
\begin{equation}
F_{ij}=\sum_{\alpha\in exp}\sum_{\beta\in pix}\left(\frac{t_{tot}}{d^{2}}\right)^{2}\frac{1}{\sigma_{\alpha\beta}^{2}}\frac{\partial(fP{}_{\alpha\beta})}{\partial p_{i}}\frac{\partial(fP{}_{\alpha\beta})}{\partial p_{j}},\label{eq:10}
\end{equation}
where $\sigma_{\alpha\beta}^{2}$ is the variance of the source signal
in each pixel introduced in equation \ref{eq:20}, and $P_{\alpha\beta}(x_{0},y_{0})$ is the value of
the ePSF centered at $x_{0}$ and $y_{0}$ at pixel position $\beta$ in exposure $\alpha$.
The two summations can be thought of as one
summation over a fine supergrid that interlaces all the pixel positions of all dither exposures.

\begin{figure}[!t]
\center
\includegraphics[scale=0.4]{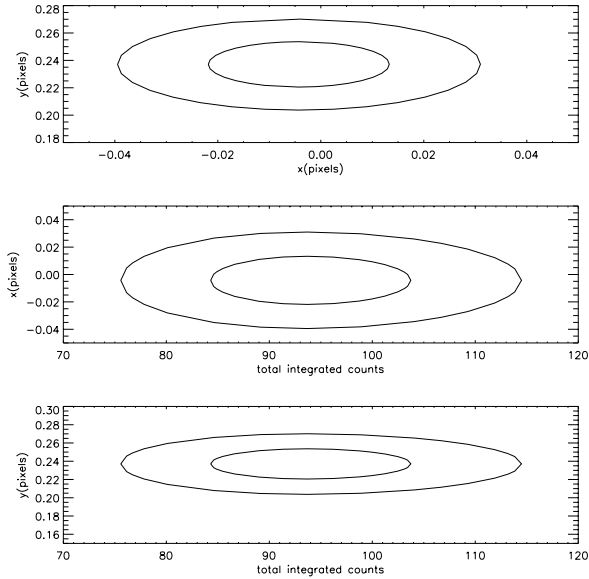}
\caption{{\footnotesize Simulated error contours for permutations of $f$, $x_0$, and $y_0$ for an extreme configuration with a pixel size of $5 l_{foc}\lambda/D$ and zero diffusion exposing with a perfect $2 \times 2$ dither pattern. The contours are similar to those expected from a Gaussian distribution. }}
\label{fig:99}
\end{figure}

To check the validity of the Gaussian assumptions inherent in using the FIM for estimating uncertainties, we perform an explicit calculation of the likelihood surface for a realization of an extreme configuration. Figure \ref{fig:99}
shows the calculated error contours using a pixel size of $5 l_{foc}\lambda/D$ and zero diffusion exposing with a perfect $2 \times 2$ dither pattern.  The source has an expected total integrated counts of 100 and is centered in the middle of a pixel and between two pixels in the two $x$-dithers and a quarter-pixel offset for the $y$-dithers.  Poisson statistics are applied in the calculation of the likelihood. We confirm that in this extreme case the confidence regions  for the three permutations of the $f$, $x_0$, and $y_0$ parameters are close to elliptical and that our Gaussian assumptions are reasonable.

\section{STATISTICAL BEHAVIOR OF PHOTOMETRIC UNCERTAINTIES}
\label{uncertainties:sec}
We now turn our attention to the photometric uncertainties of the ensemble of point sources randomly distributed on the pixel grid.
This is of interest to multiplexed surveys where multiple objects lie within the imager field of view.  
As seen in the previous section, PSF-fit parameter uncertainties of a single source
depend on the supergrid (the interlaced pixel grids of all the pointings) and the source centroid position relative to the grid. Averaging
over randomly positioned sources, the difference in
parameter estimation can only be due to differing dither patterns and number
of dithers.  Our interest is to determine
which pixel scales and dither patterns work well for the set of objects as a whole rather than for an individual object.

The statistics we consider are the mean and standard deviations of the parameter uncertainties.  
Clearly low mean uncertainties are beneficial for any survey. However, the importance of the standard deviation depends on the science and the survey strategy.  Multiepoch observations of quiescent objects can be analyzed with all the dithers of all visits.
 A large standard deviation produces large 
variations in uncertainty between objects and between different epochs of the same object.  
The realized uncertainty depends on the subpixel position of the object meaning that
 there is a nontrivial efficiency window-function over the sky. 
  The standard deviation must be accounted for in calculating the multiplex efficiency of observing multiple
   objects in a pointing.  This is particularly important for time-variable objects; for example the fit for distance can depend
strongly on which visits happen to have extremely inaccurate measurements.

This section is organized as follows: We first introduce the hardware setups and dither patterns considered in the analysis, simulation details, notation, and units.
First-order results are derived from analytical expressions for the
Fisher elements in Equation~\ref{eq:10}. We then show numerical results
from simulations where we scan over pixel sizes and detector diffusions for different dither patterns.

\subsection{Analysis Overview and Technical Details}
\label{overviewdetail:sec}
We calculate distributions of photometric uncertainties for a range of dither strategies, hardware properties, and priors on the position of the source of interest.
The dithering strategies include $2 \times 2$ and $3 \times 3$ dither pointings (labeled by d2 and d3)
each with three different patterns.  One pattern has completely
random pointings (random dithering labeled with the subscript R), the second uses precise dither steps to give a uniformly spaced
supergrid (perfect dithering labeled with the subscript P), the third attempts perfect dithering but with a random Gaussian pointing uncertainty
(Gaussian dithering labeled with the subscript G).
Hardware scenarios cover pixel scales $a$ ranging between $1$ and $5$ and detector diffusions $\sigma_{diff}$  ranging between $0$ and $3$,
both in 0.2 steps in units of the diffraction scale $l_{foc}\lambda/D$.
These ranges include scenarios where the ePSF is dominated by
an Airy disk, Gaussian, and top-hat functions.   

The flux uncertainty of a single observation is expressed as
$\sigma(f)=\sqrt{(F^{-1})_{ff}}$; position uncertainties affect the final flux uncertainty as $F$ is not diagonal.
Cases where the source position is known independently are given by $\sigma(f_{0})=\sqrt{(F_{ff})^{-1}}$.
Fitting for the flux yields different uncertainties compared with when both flux and position are fit.

The square root of the area of the $x$--$y$ error ellipse, as we here denote  $\sigma(c)$, is used to represent the centroid position uncertainty. In terms of the Fisher matrix,  $\sigma(c)$ is then given as the fourth root of the determinant of the $x$--$y$ submatrix of the full covariance matrix $F^{-1}$.

Uncertainty results are given relative to that of the hardware choice of $\sigma_{diff}=1$, $a=2$, where diffraction, diffusion, and pixel response have similar contributions to the ePSF.

As described in the introduction to the section, we are interested in the average performance from a given dither pattern and pixel scale, rather than from a particular pointing. The distribution from which we can find characteristics like the average and variance is found by generating  $10^{4}$ random realizations for the initial
pointing for each choice of pixel scale, diffusion, and dithering strategy. To simplify the notation, we express the parameter set of interest in the vector $\mathbf{w} = \{\sigma(f), \sigma(c)\}$, the
 average of a particular parameter $w$ is denoted by $\left\langle w\right\rangle $, and the square root of the variance
$stdev{\left(w\right)}\equiv\sqrt{\left\langle [\sigma(w)-\left\langle \sigma(w)\right\rangle ]^{2}\right\rangle }$.

\subsection{Analytical Average of the Fisher Elements}
\label{sub:meanFapp}

Before we consider the results from simulations, we now present first-order calculations for  $\left\langle \sigma(f)\right\rangle $ 
and $\left\langle \sigma(c)\right\rangle $. 
From these, we clearly see the dependence on different pixel scales under different noise limits. 
The results help clarify some of the general features in the simulation plots presented in the next section.

 The first-order results are defined to describe $\left\langle \sigma(p_{i})\right\rangle $ in the limit where 
 the variance of the terms in equation~\ref{eq:10} is small. In that limit $\left\langle \sigma(p_{i})\right\rangle $ can be well approximated by
 $1/\sqrt{ \left\langle F_{p_{i}p_{i}}\right\rangle }$, reducing the problem to a calculation of  $\left\langle F\right\rangle $, which turns out to have
 a relative simple analytical form.  The calculations in the Appendix show that $\left\langle F\right\rangle $
is diagonal, symmetric in $x$ and $y$, and completely independent
of the dither pattern. 
 The only two unique nonzero terms (out of nine) 
in $\left\langle F\right\rangle $
are
\begin{equation}
\left\langle F_{ff}\right\rangle =\frac{a^{2}t^{2}}{d^{2}}\int_{-\infty}^{+\infty}\frac{P^{2}}{\sigma^{2}}ds\label{eq:30}\end{equation}
and
\begin{equation}
\left\langle F_{xx}\right\rangle =\frac{a^{2}f^{2}t^{2}}{d^{2}}\int_{-\infty}^{+\infty}\frac{P_{,x}^{2}}{\sigma^{2}}ds,\label{eq:40}
\end{equation}
where $\left\langle F_{yy}\right\rangle =\left\langle F_{xx}\right\rangle $
by symmetry. Here, $P_{,i}$ denotes the derivative of $P$
with respect to parameter $i$. In this limit,   $\left\langle \sigma(c)\right\rangle $ equals 
 $\left\langle F_{xx}\right\rangle ^{-1/2}$.
In general, in an experiment it is of interest to increase the value of $\left\langle F_{p_{i}p_{i}}\right\rangle$, 
since this will reduce the overall fitting uncertainty. 

We now discuss first-order scalings of fit uncertainties in different noise limits.
Writing out the noise terms in Equations~\ref{eq:30}
and \ref{eq:40} shows  that $\left\langle F_{ff}\right\rangle$ and $\left\langle F_{xx}\right\rangle$ in the sky- and source-noise limits
are both 
independent  of the number of dithers. Furthermore, the term $\left\langle F_{ff}\right\rangle $ in the 
source-noise limit ($\left\langle F_{ff}\right\rangle _{source}$) is  independent of the shape of $P$: i.e., \ $I(q)$, $\sigma_{diff}$,
and $a$.

In general, the integrals in equations~\ref{eq:30} and \ref{eq:40}
must be calculated numerically, but  analytic results
exist in some limits. Approximating the Airy function by a Gaussian with width
$\sigma_{tel}$, the function $P$ in the well-sampled limit
is described by a Gaussian. Then $\int P^{2}ds \propto1/\sigma_{con}^{2}$ and $\int {P_{,x}}^{2}ds \propto 1/\sigma_{con}^{4}$,
where $\sigma_{con}^{2}=\sigma_{diff}^{2}+\sigma_{tel}^{2}$ is the
total width of the convolved function $P$.   Then $\left\langle F_{ff}\right\rangle _{sky}^{-1/2}$
scales as $\sigma_{con} \sqrt{n_{s}/(tl_{foc}^2)}$, $\left\langle F_{ff}\right\rangle _{source}^{-1/2}$
scales as $\sqrt{f/t}$, $\left\langle F_{xx}\right\rangle _{sky}^{-1/2}$
 scales as $\sigma_{con}^{2} \sqrt{n_{s}/(tf^2l_{foc}^2)}$ and $\left\langle F_{xx}\right\rangle _{source}^{-1/2}$
scales as $\sigma_{con}/\sqrt{ft}$.
When dominated by the pixel, $\int P^{2}ds=1/a^{2}$, and ${P_{,x}}$ is ill-defined:
$\left\langle F_{ff}\right\rangle _{sky}^{-1/2}$ scales as
$a/l_{foc}$.

\subsection{Simulation of Different Dither Strategies}
\label{DifferentPointings:sec}
This subsection presents numerical calculations of the photometric uncertainty distributions due to different dither patterns and pixel scales.
Perfect dithering has been shown to be optimal
for image reconstruction \citep{1999PASP..111..227L}, but imposes pointing requirements
on the telescope.  On the other hand, a random dithering imposes no pointing requirements on the
telescope, simplifying the mission design.   We present these dither patterns as follows:
The uncertainty distributions for the random dither patterns are shown first in \S\ref{random:sec}.
The differences between prefect and random dither patterns are given in \S\ref{randomvperfect:sec}.
Pointing requirements are drawn from the analysis of \S\ref{sub:GD-section}, whose dither pattern includes a pointing
error when attempting perfect dithering.

We consider both cases where the source flux and position are derived from the 
data
and where the centroid position is already known based on other data. The latter situation
approximates the photometry of a single point on a densely sampled light curve, where the star/supernova position is derived from
all other pointings. 
The read-noise-dominated regime is not considered, as increasing the number of exposures
with dithering is then clearly disfavored.
The average flux uncertainties in the source-noise-dominated limit are not presented, 
since they are neither dependent on the shape of ePSF nor on the dither pattern, as shown in \S\ref{sub:meanFapp}. 

\subsubsection{Random Dither Pattern}
\label{random:sec}
We begin by calculating the average flux and position uncertainties for random dithers in the sky- and source-noise-dominated limits, and
the fractional differences of those average uncertainties between using a $2\times2$ and a $3\times3$
dither.

\begin{figure}[!t]
\center
\includegraphics[scale=0.4]{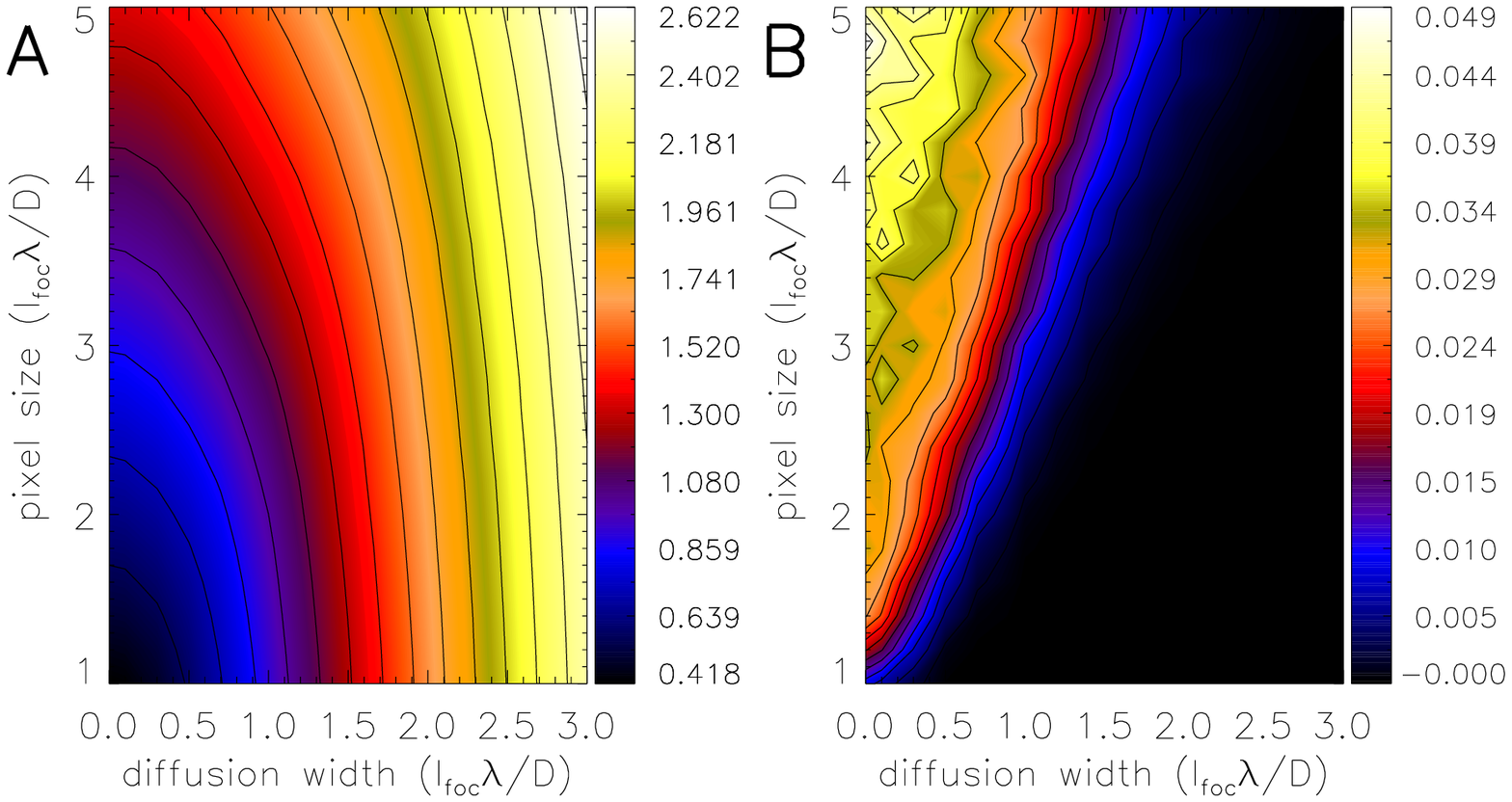}\\

\includegraphics[scale=0.4]{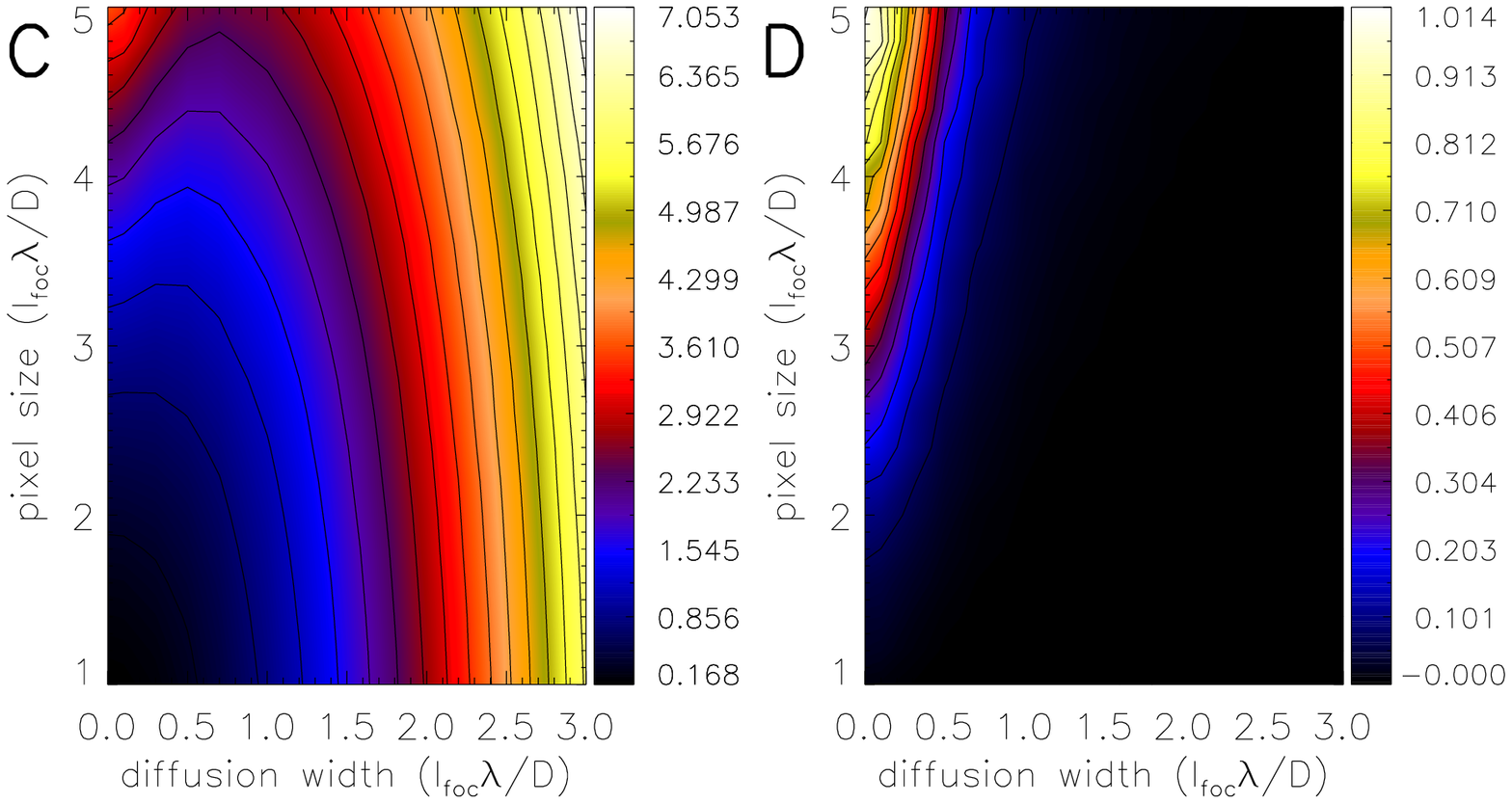}\\

\includegraphics[scale=0.4]{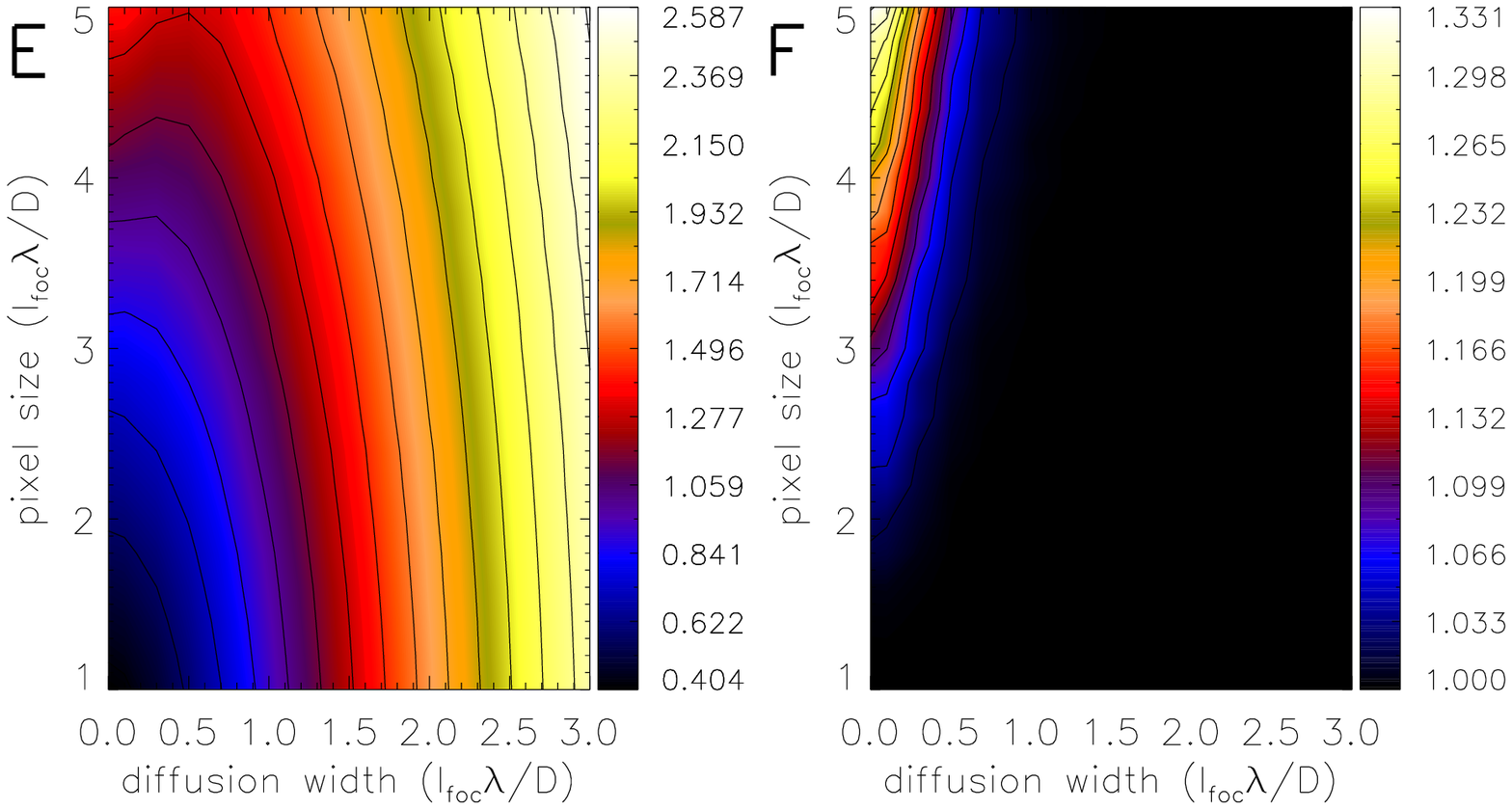}\\

\caption{{\footnotesize Average flux and position uncertainties for the case of random dither patterns.  \emph{Left}: Uncertainties for $2\times2$ dithering: $\left\langle \sigma(f)_{R}\right\rangle _{sky}$ (\emph{top}), $\left\langle \sigma(c)_{R}\right\rangle _{sky}$ (\emph{middle}),
and $\left\langle \sigma(c)_{R}\right\rangle _{source}$ (\emph{bottom}).
\emph{Right}: Corresponding relative difference in the average uncertainties for the $2\times2$ and $3\times3$ cases as $\left\langle w_{R}\right\rangle _{d2}/\left\langle w_{R}\right\rangle _{d3}-1$.  Each plot in the left column is scaled with its value at $\sigma_{diff}=1$ and $a=2$.}}
\label{fig:10}
\end{figure}

The plots on the left of  Figure~\ref{fig:10} show (\emph{from top to bottom}) $\left\langle \sigma(f)_{R}\right\rangle _{sky}$,
$\left\langle \sigma(c)_{R}\right\rangle _{sky}$, and $\left\langle \sigma(c)_{R}\right\rangle _{source}$ 
as functions of the variables $\sigma_{diff}$ and $a$ for a $2\times2$ dither pattern.  
The right column shows the relative difference between using $2\times2$ and $3\times3$ dithering for the same parameter set and 
order of noise as  $\left\langle w_{R}\right\rangle _{d2}/\left\langle w_{R}\right\rangle _{d3}-1$.

The plots in the left column of Figure~\ref{fig:10}  show that in the sky- and source-noise limit,
the average uncertainties of all the parameters ($\left\langle w_{R}\right\rangle$) 
becomes smaller as the pixel size decreases, for a fixed detector diffusion $\sigma_{diff}$.
On average, the fit for the PSF photometry becomes better as the pixel contributes less to the ePSF.
This is in agreement with intuition and the calculations in \S\ref{sub:meanFapp}.

The average flux uncertainties
$\left\langle \sigma(f)_{R}\right\rangle$ in the sky-noise-dominated case are proportional to the size of the ePSF.
(Recall that the source-noise-dominated case is not shown, since the average flux uncertainties are then only weakly
dependent on the ePSF.)
On the other hand, the average position uncertainties $\left\langle \sigma(c)_{R}\right\rangle $
depend on the shape in addition to the size of the ePSF.  The position determination depends on the derivatives of the ePSF, which 
has sharp features when dominated by the pixel. Intuitively, such a degradation
in the position determination is expected, since centroid information within a single image is lost when the source signal lands
in only one pixel.  Then dithering can only localize the centroid down to the scale of the supergrid spacings.

Indeed, the comparison of the $2\times2$ and $3 \times 3$ dithers in the right column of Figure~\ref{fig:10} shows that increased dithering 
reduces uncertainties only in regions where the pixel dominates the ePSF.
For the well-sampled region we see 
no measurable difference as expected from the first-order estimates. 

An interesting conclusion drawn from these calculations is that when fitting for the centroid position, degrading the width of the PSF with a Gaussian blur (say, by
defocusing the telescope) can produce improved signal-to-noise ratio, despite the increase in sky noise.   For pixel sizes above a minimum threshold, the optimal diffusion
width is nonzero.

\begin{figure}[!t]
\center
\includegraphics[scale=0.40]{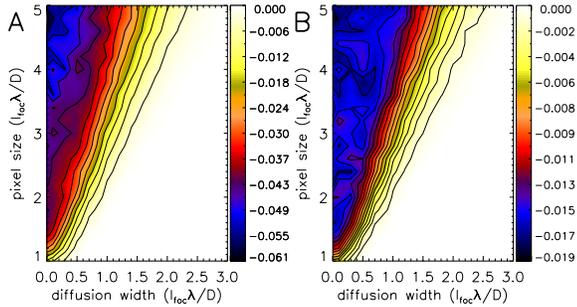}
\caption{{\footnotesize Relative difference between the average flux uncertainties when the position of the source is independently
known and when that position is determined in the PSF-photometry fit,
$\left\langle \sigma(f_{0})_{R}\right\rangle /\left\langle \sigma(f)_{R}\right\rangle -1$.
\emph{Left}:  $2\times2$ dither. \emph{Right}: $3\times3$
dither. }}
\label{fig:205}
\end{figure}

When the position of the centroid is known the flux uncertainties decrease.  The comparison between the relative difference in the average
flux uncertainty when and when not fitting for the position
is shown in Figure~\ref{fig:205}, which shows a plot of  $\left\langle \sigma(f_{0})_{R}\right\rangle /\left\langle \sigma(f)_{R}\right\rangle - 1$.
The differences are at the $1-5$\% level and are only
appreciable when the pixel dominates the ePSF.

\begin{figure}[!t]
\center
\includegraphics[scale=0.40]{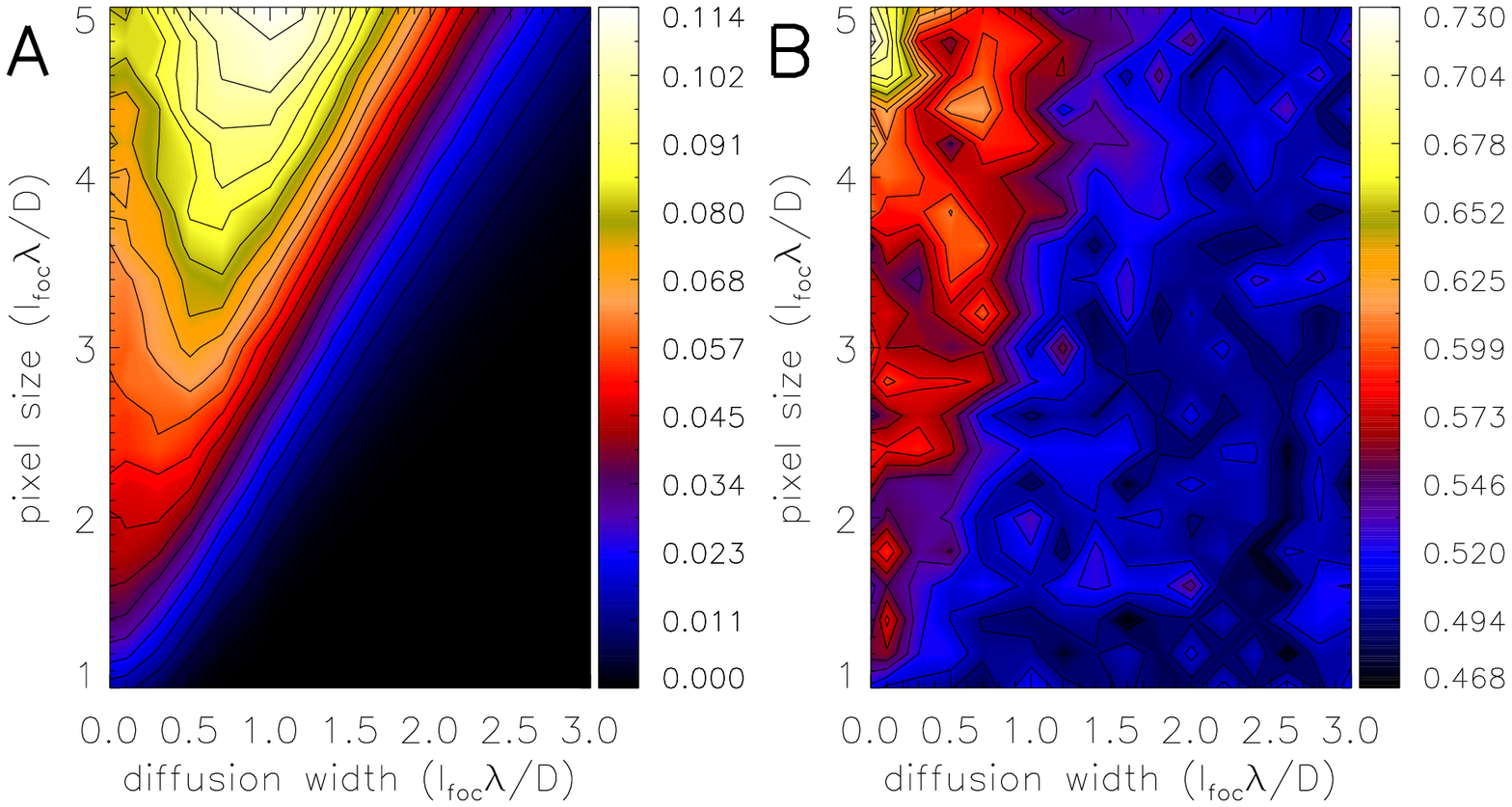}

\includegraphics[scale=0.40]{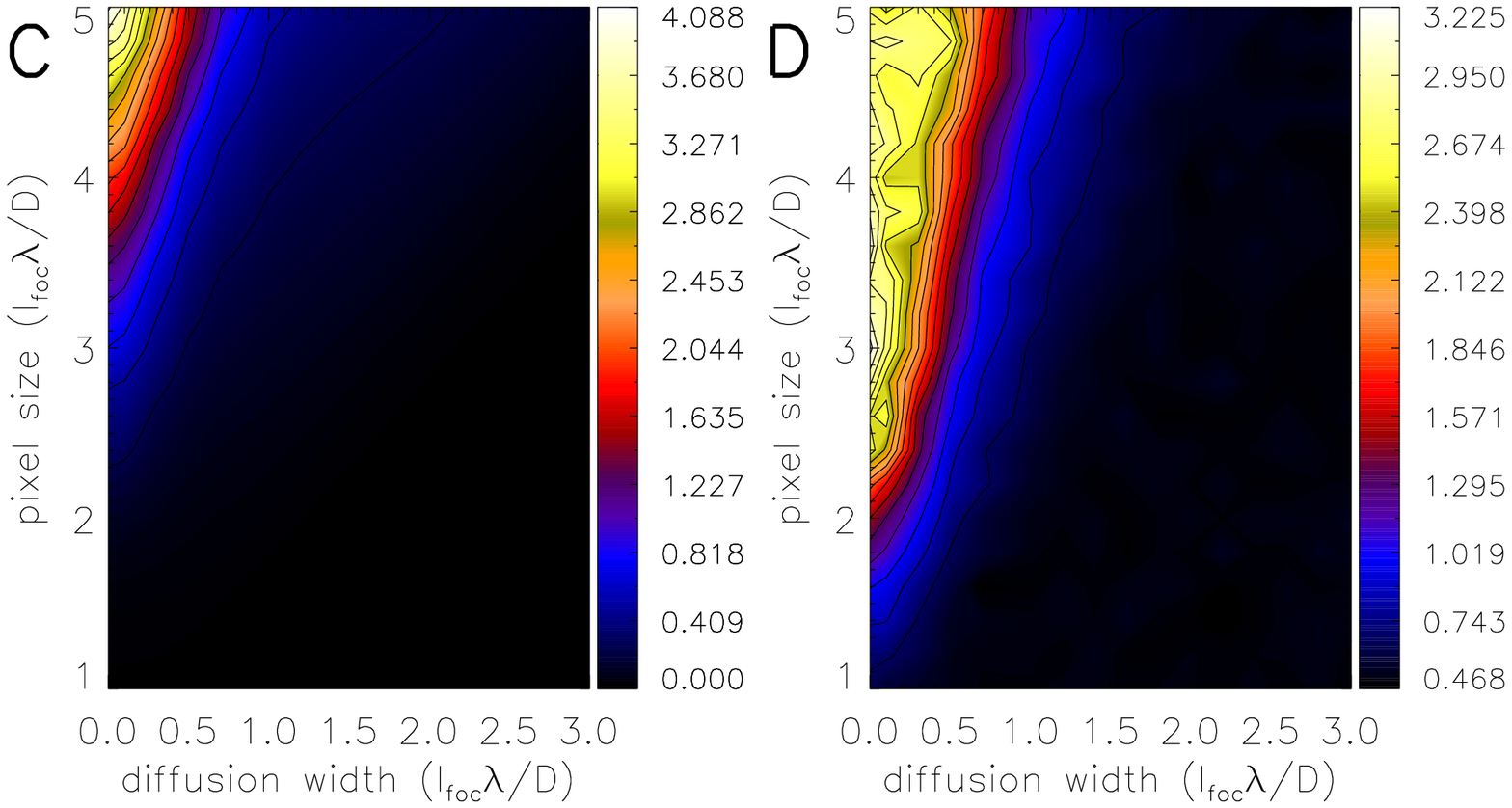}

\includegraphics[scale=0.40]{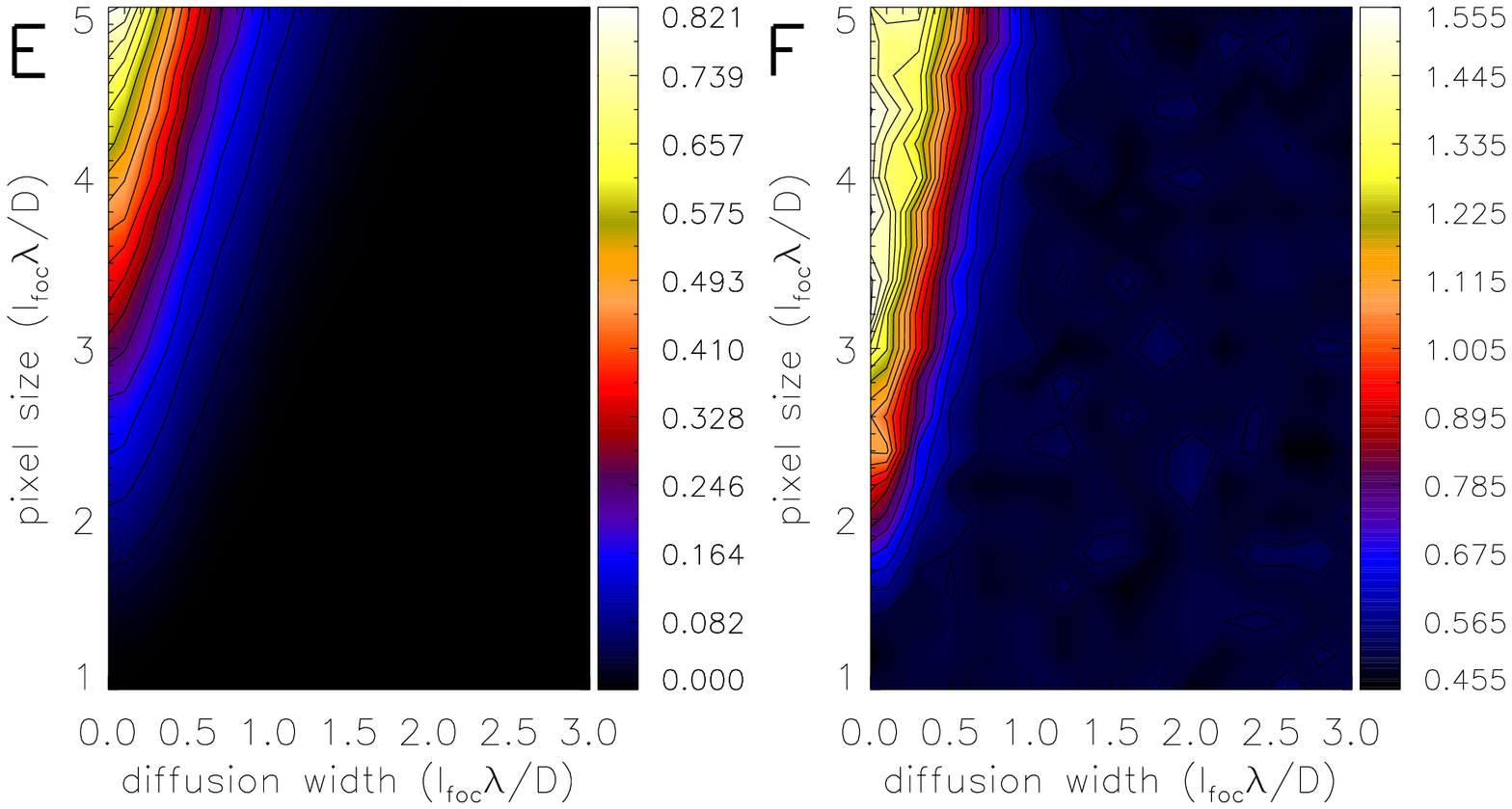}

\caption{{\footnotesize Square root of the variance of the flux and position uncertainty distributions for the case of random dither patterns.  \emph{Left}:  $stdev{\left(w\right)}$ for the $2\times2$ dither for 
 $stdev{\left(\sigma(f)_{R}\right)}_{sky}$ (\emph{top}), $stdev{\left(\sigma(c)_{R}\right)}_{sky}$ (\emph{middle}),
and $stdev{\left(\sigma(c)_{R}\right)}_{source}$ (\emph{bottom}). \emph{Right}: Corresponding relative difference in $stdev{\left(w\right)}$ for the $2\times2$ and $3\times3$ cases as $stdev{\left(w_{R}\right)}_{ d2}/ stdev{\left(w_{R}\right)}_{d3}-1$. The plots in the left column are individually scaled with the same factor as in Fig. \ref{fig:10}.}}
\label{fig:201}

\end{figure}

Having discussed the average, we now turn to the second moment of the uncertainty distributions through the
statistic
$stdev{\left(w\right)}$.
Figure~\ref{fig:201} %
shows $stdev{\left(w\right)}$ for flux and uncertainty distributions when using  
$2\times2$ and $3\times3$ random dither patterns.

The variance is small when the PSF is well sampled and increases as the pixel dominates the ePSF.  While the variance of the average position uncertainty
directly corresponds to the pixel contribution to the ePSF size, the variance for the average flux uncertainty also depends on the shape of the PSF; the
variance does not increase as quickly with larger pixels when the underlying PSF is an Airy function as opposed to a Gaussian.
This is seen in the   $V$-shaped contours in $stdev{\left(\sigma(f)\right)}$.

The ratio between the average and width of the uncertainty distributions is found by comparing Figures~\ref{fig:10}
and \ref{fig:201}. 
The flux-uncertainty distribution is well localized within  $\left\langle \sigma(f)_{R}\right\rangle/stdev{\left(\sigma(f)_{R}\right)} \sim 0.08$ for $2\times2$ dithering and $0.05$ for $3\times3$. On the other hand, the
position uncertainty is broad, as $\left\langle \sigma(c)_{R}\right\rangle$ and $stdev{\left(\sigma(c)_{R}\right)}$ are comparable in size.

\begin{figure}[!t]
\center
\includegraphics[scale=0.40]{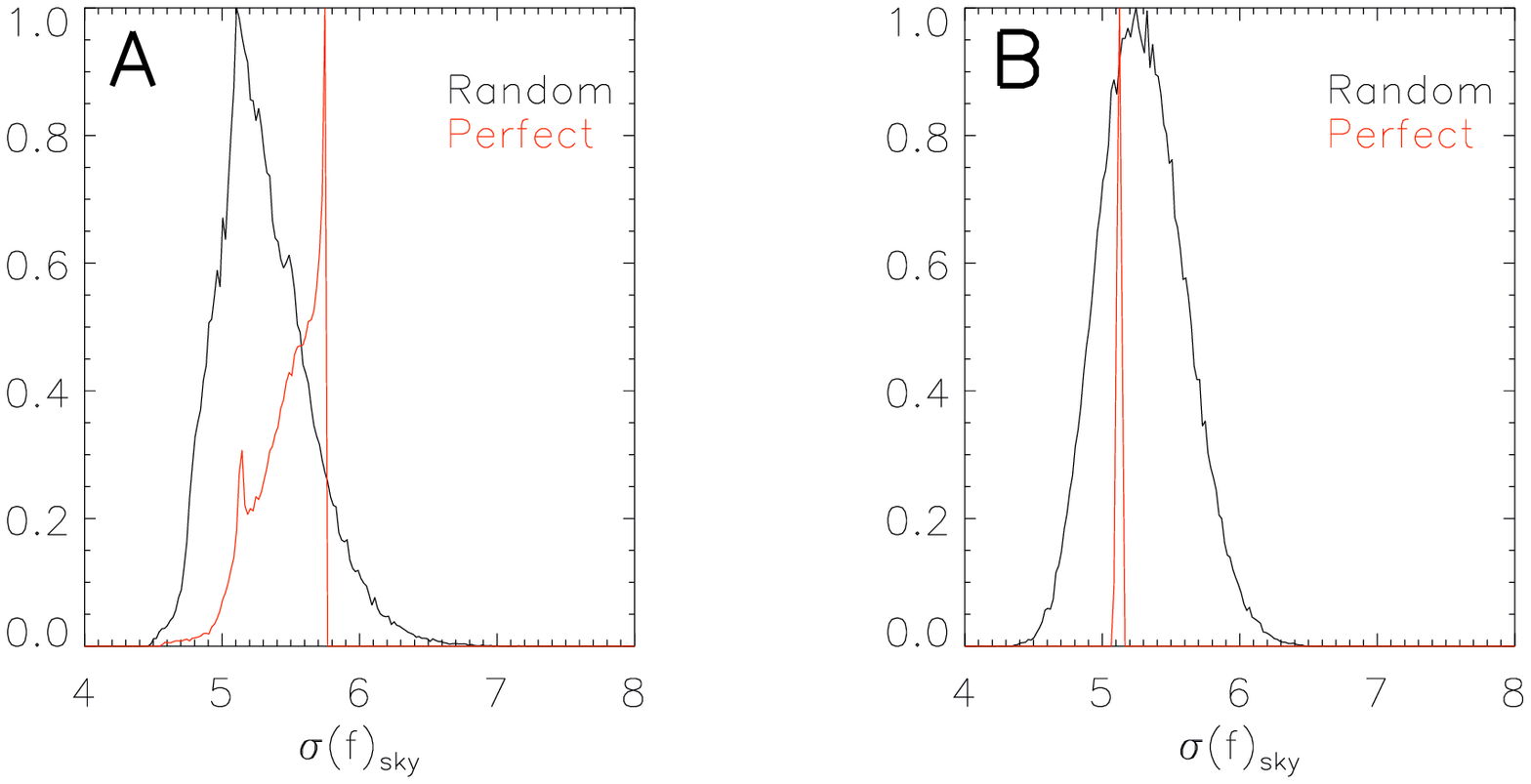}
\includegraphics[scale=0.40]{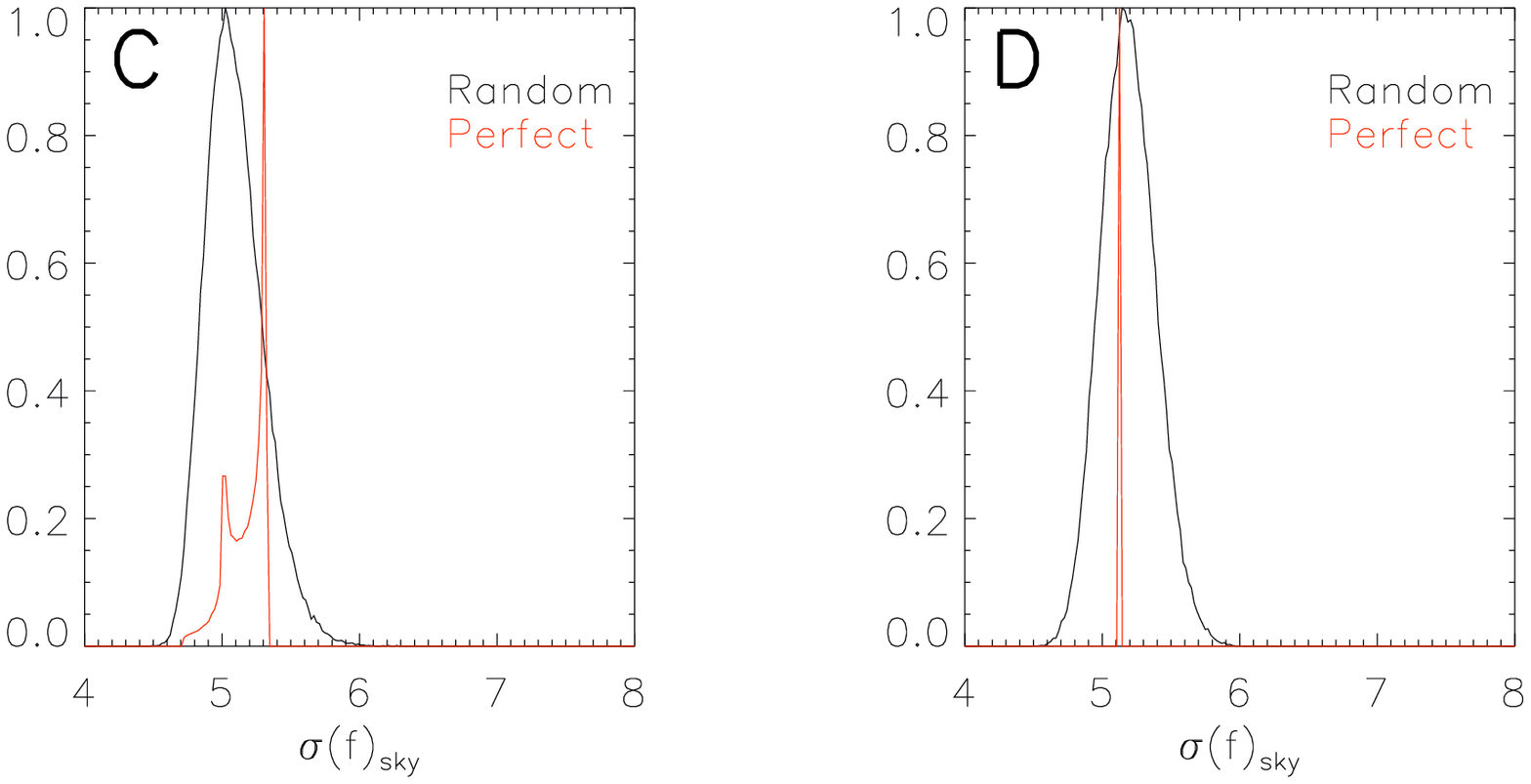}
\caption{{\footnotesize Distributions of $\sigma(f)_{sky}$ for the two
points in the diffusion--pixel-scale space: $(0,\ 4.1)$ (\emph{left}) and $(0.83,\ 3.2)$ (\emph{right}). The ePSF is pixel-dominated
for the first point and has comparable diffraction, diffusion, and pixel contributions for the second point. The red
curve is the distribution when using a perfect dither pattern and the black curve 
is when using a random pattern. \emph{Top}: $2\times2$ dithering. \emph{Bottom}: $3\times 3$ dithering. The histograms are based
on $10^{5}$ randomly realized pointings.}}
\label{fig:200}
\end{figure}

The right column in Figure \ref{fig:201} shows that increasing the number of dithers  reduces $stdev{\left(w\right)}$; the
width using a $2\times2$ dither grid is at least 50\% wider than the width when using a $3\times3$ grid.
The smaller dispersion is readily apparent in Figure~\ref{fig:200}, which shows the full distribution of $\sigma(f)_{sky}$ for two sample points in
the $\sigma_{diff}$--$a$ parameter space.

\subsubsection{Random Versus Perfect Dither Pattern}
\label{randomvperfect:sec}
\begin{figure}[!t]
\center
\includegraphics[scale=0.40]{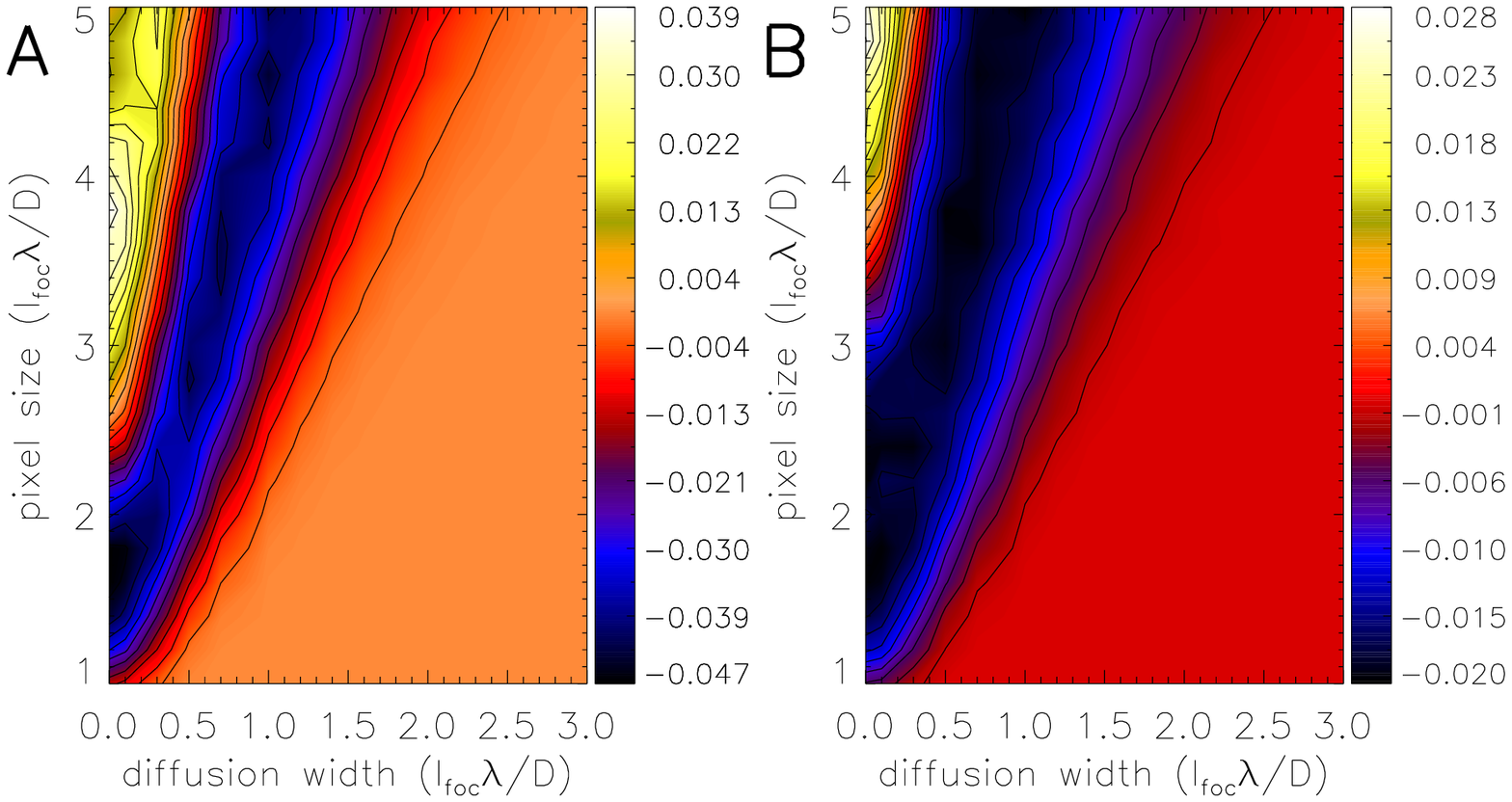}
\includegraphics[scale=0.40]{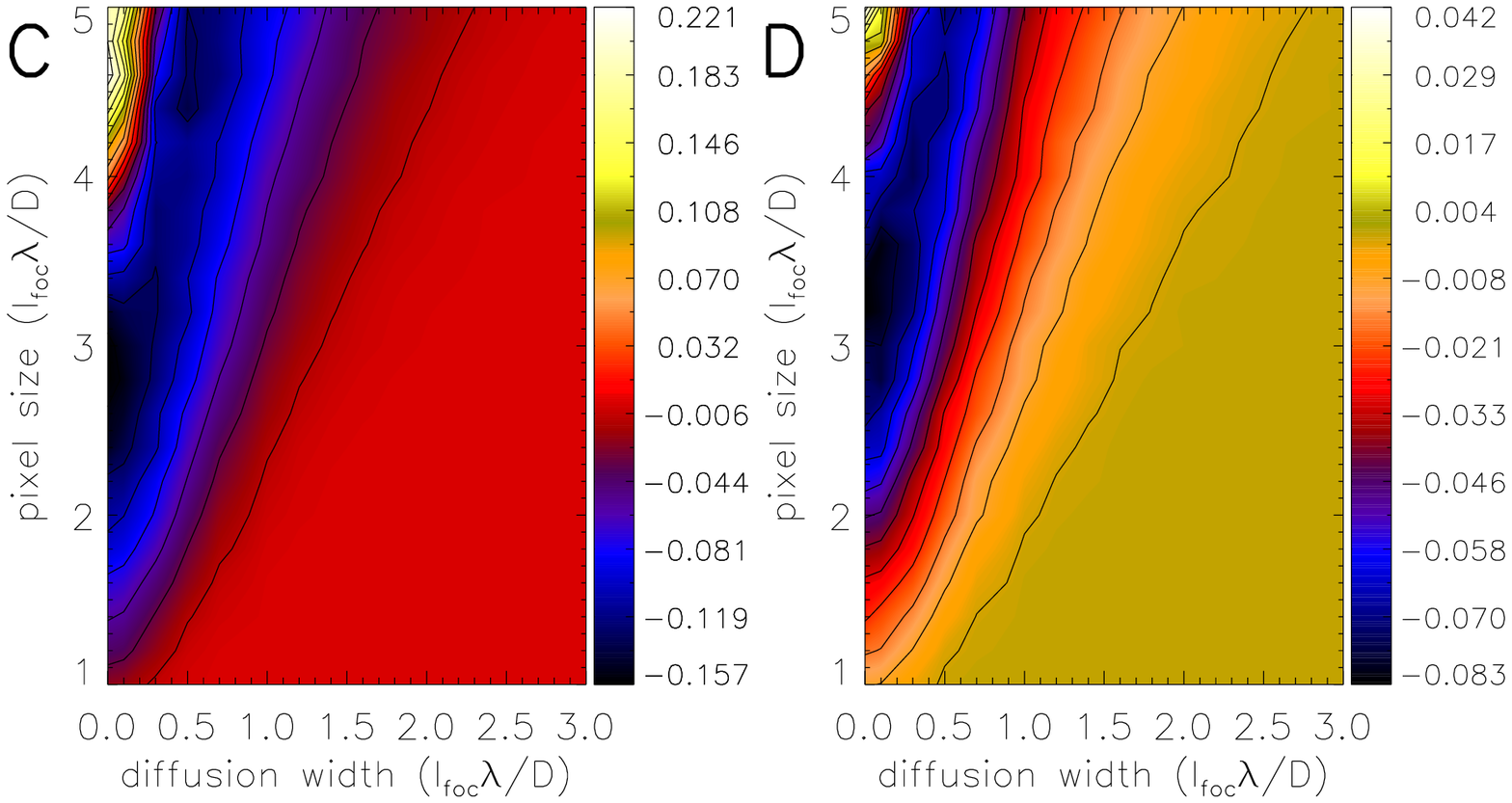}
\includegraphics[scale=0.40]{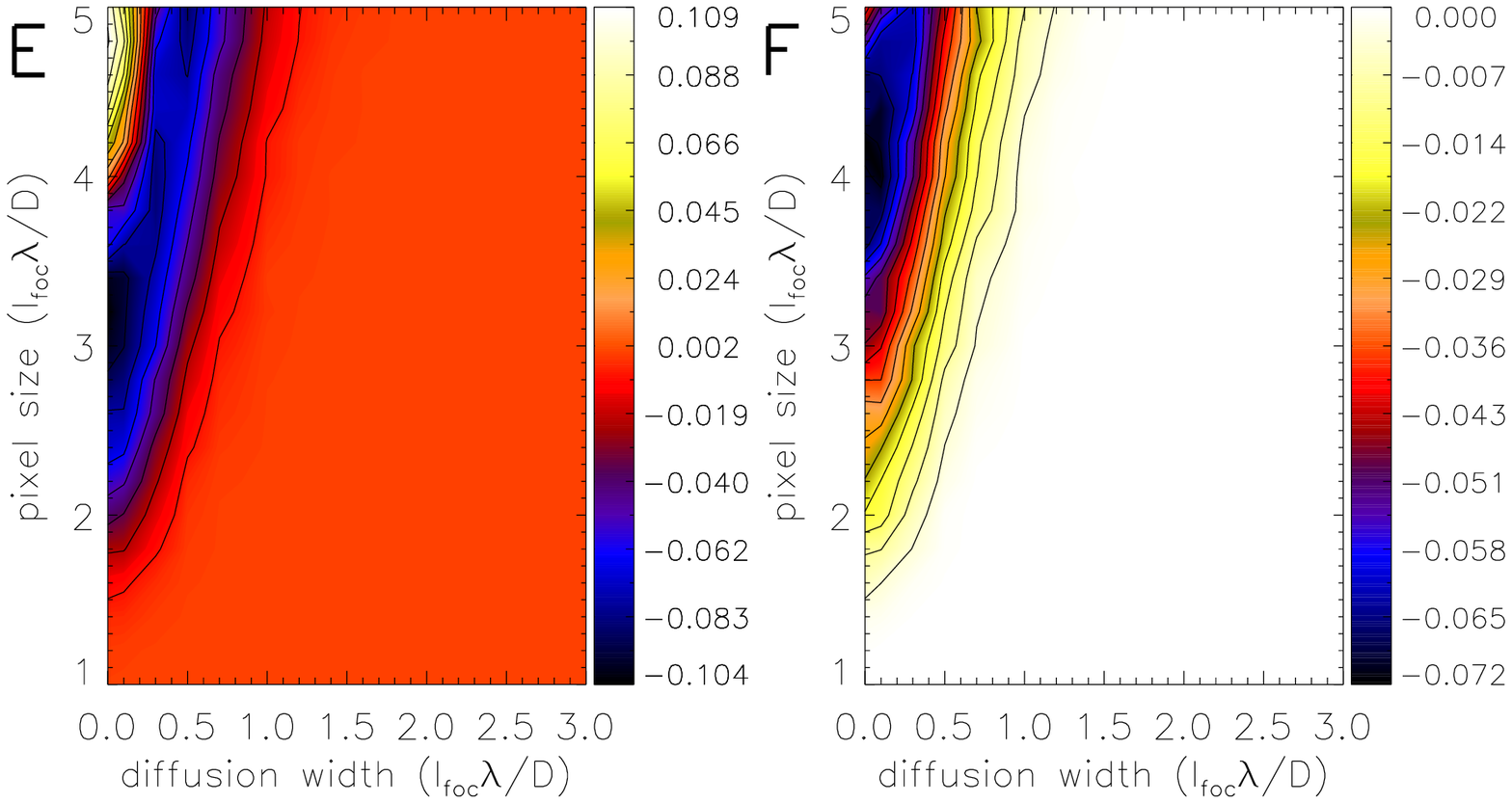}
\caption{{\footnotesize Relative difference between  perfect and random dither patterns in the average parameter uncertainties given by
$\left\langle w_{P}\right\rangle /\left\langle w_{R}\right\rangle -1$.  The three rows from the top correspond to the
flux uncertainties in the sky-noise limit and position uncertainties
in the sky-
and source-noise limits:  $\sigma(f)_{sky}$ (\emph{top}), $\sigma(c)_{sky}$ (\emph{middle}), 
and $\sigma(c)_{source}$ (\emph{bottom}). \emph{Left}: $2\times2$ dither. \emph{Right}: $3\times3$ dither.}}
\label{fig:20}
\end{figure}

The relative differences between the average parameter uncertainties when using perfect versus random dithers,  $\left\langle w_{P}\right\rangle /\left\langle w_{R}\right\rangle -1$,
are shown in Figure~\ref{fig:20}.
The dither patterns are equivalent when the PSF is well sampled, as expected from the discussion in \S\ref{sub:meanFapp}.
When not well sampled, the difference in $\sigma(f)$ when sky-noise-dominated
is higher by up to about $2-5$\% depending on the number dithers used, while
for the centroid position the difference can be up to almost
$25$\%. However, the most interesting observation is that in some regions perfect
dithering has lower average uncertainties, whereas in other regions, random dithering gives lower averages.

\begin{figure}[!t]
\center
\includegraphics[scale=0.40]{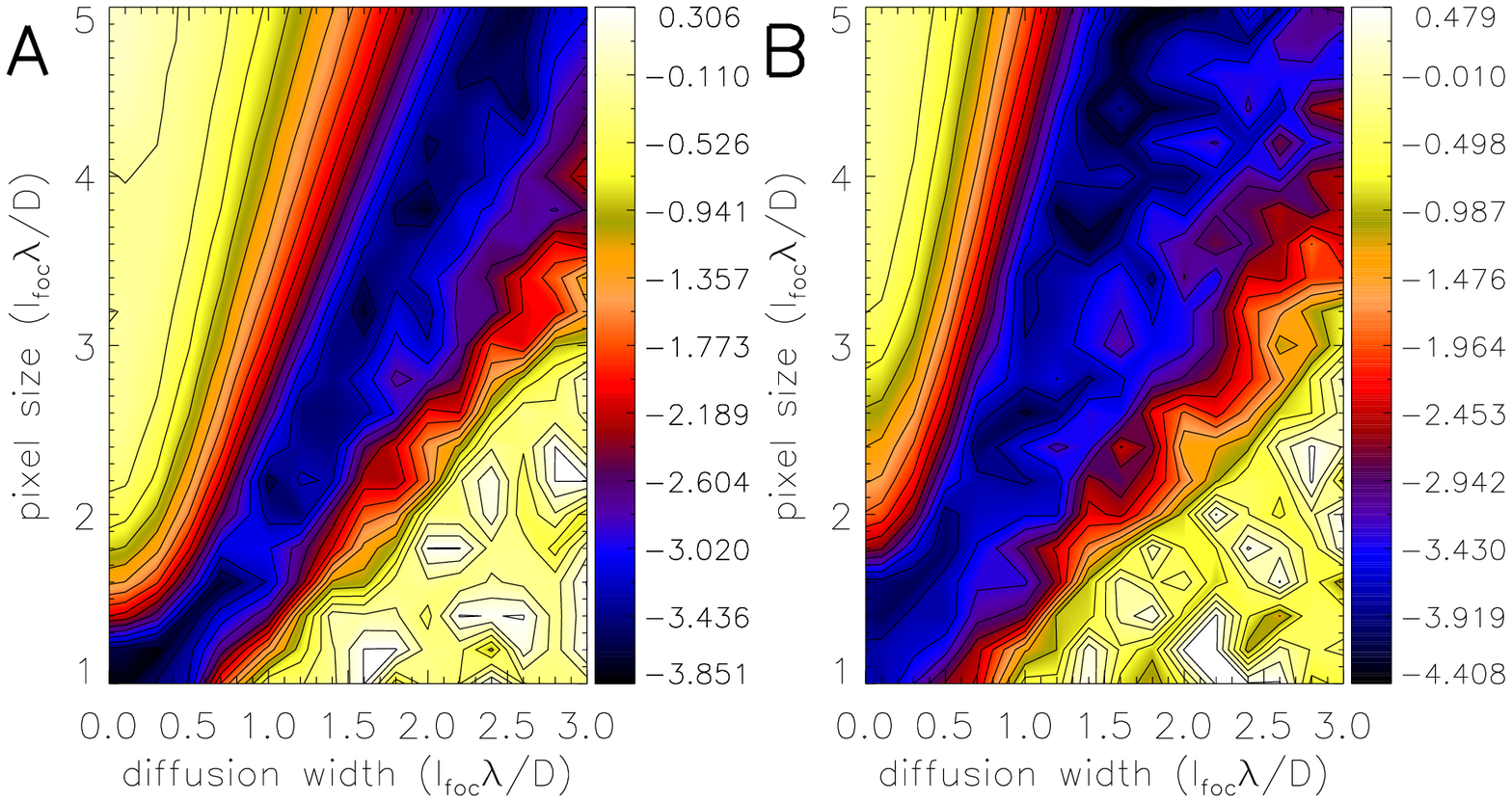}
\includegraphics[scale=0.40]{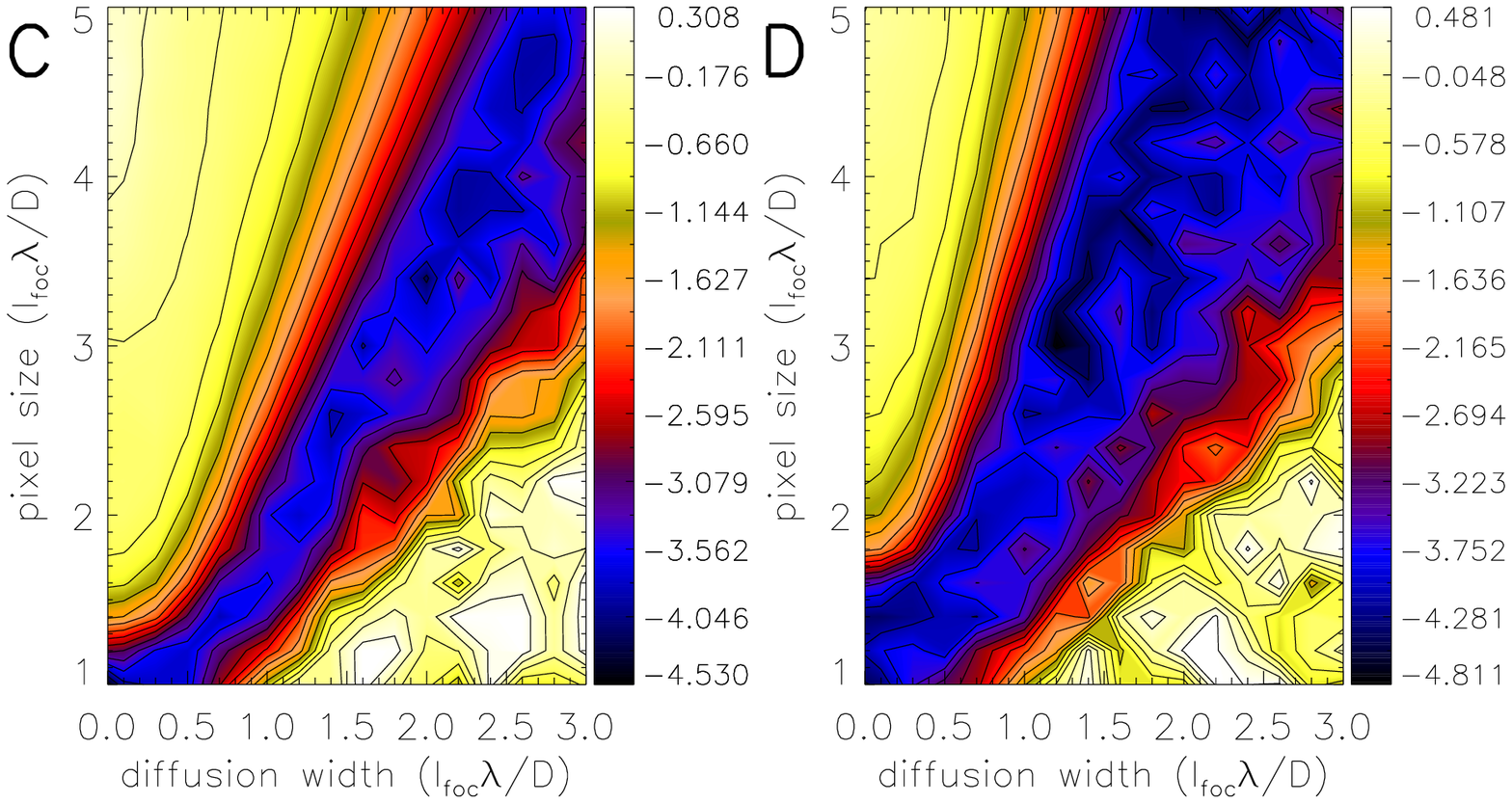}
\includegraphics[scale=0.40]{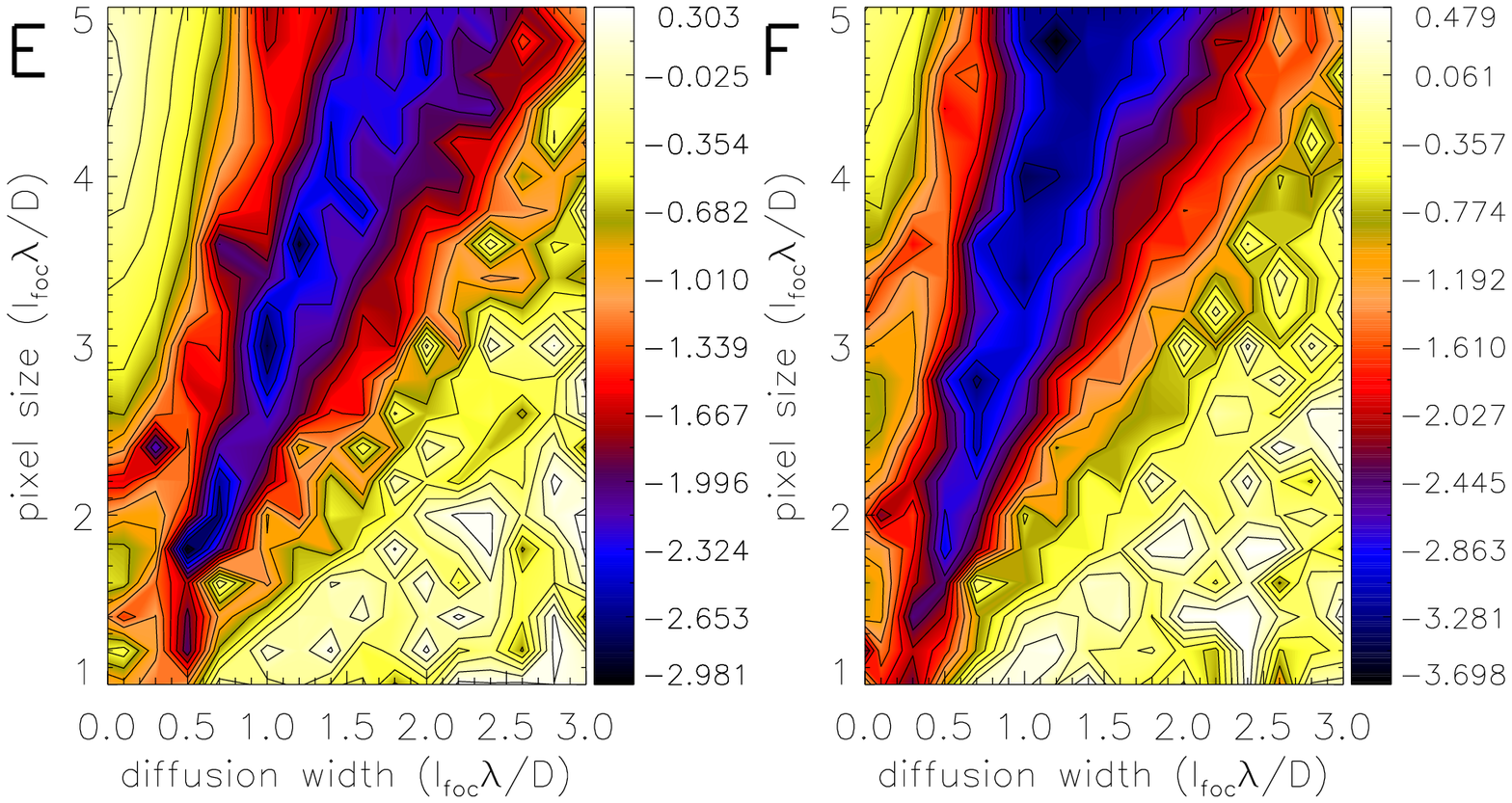}
\caption{{\footnotesize Comparison between the square root of the variances from perfect and random dither patterns.  The three rows from the top correspond to the
flux uncertainties in the sky-noise limit and position uncertainties
in the sky-
and source-noise limits:  $\log_{10}{\left(stdev{\left(\sigma(f)_{P}\right)}/stdev{\left(\sigma(f)_{R}\right)}\right)}_{sky}$ (\emph{top}), $\log_{10}{\left(stdev{\left(\sigma(c)_{P}\right)}/stdev{\left(\sigma(c)_{R}\right)}\right)}_{sky}$ (\emph{middle})
and  $\log_{10}{\left(stdev{\left(\sigma(c)_{P}\right)}/stdev{\left(\sigma(c)_{R}\right)}\right)}_{source}$ (\emph{bottom}).
\emph{Left}: $2\times2$ dither. \emph{Right}: $3\times3$ dither.}}
\label{fig:22}
\end{figure}

The differences between using different patterns are more pronouncedly manifest in the variances.
Figure \ref{fig:22} shows the logarithm of the ratio between $stdev{\left(w_{P}\right)}$ 
and $stdev{\left(w_{R}\right)}$; the width of the $w$ distributions are reduced
by up to several orders of magnitude by shifting from random 
to perfect dithering.

The full distributions of the parameter uncertainties exhibit the non-Gaussian behavior when there are coarse pixel scales and/or
perfect dithering. Distributions of flux uncertainties $\sigma(f)_{sky}$ over random initial pointings are shown in Figure~\ref{fig:200}.
Shown are two detector-diffusion/pixel-scale pairs with similar average flux
uncertainty: $\sigma_{diff}=0$ and $a=4.1$, for which ePSF is strongly dominated by the pixel,
and $\sigma_{diff}=0.83$ and $a=3.2$, which has comparable contributions from each source of blur.
For $0.83-3.2$, the distributions resemble Gaussians and perfect dithering gives a low average flux uncertainty with little scatter.
For $0-4.1$, random dithering has lowest average flux uncertainty and asymmetric distributions; the perfect
dithering case, in particular, has sharp edges in the $\sigma(f)$ distribution with peaks at the extremes of the range
responsible for the narrower distribution.

\begin{figure*}[!t]
\center
\includegraphics[scale=0.80]{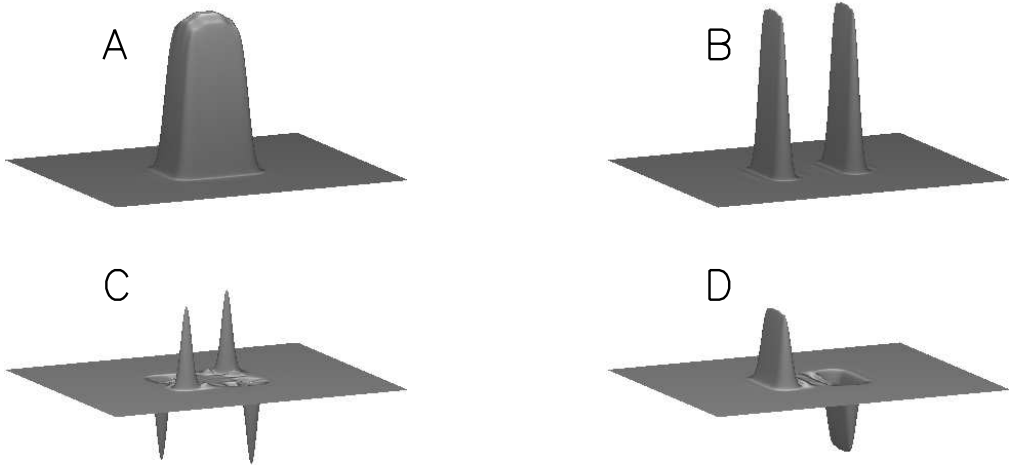}
\caption{{\footnotesize Surfaces of four different combinations of 
 $\partial (fP_{\alpha\beta})/ \partial p_{i}$ and  $\partial (fP_{\alpha\beta})/ \partial p_{j}$
 for $a=5$ and $\sigma_{diff}=0$. These functions are sampled by the supergrid in the determination of the Fisher matrix elements
using eq.~\ref{eq:10}. The combinations in terms of $(p_{i},p_{j})$ are A = $(f,f)$, B = $(x,x)$, C = $(x,y)$ and D = $(x,f)$.
The equivalent functions when the ePSF is not pixel-dominated are smooth without sharp peaks.}}
\label{fig:150}
\end{figure*}

A perfect dither pattern has only one independent pointing; all pointings are offset by fixed amounts relative to the first.
The random pattern has $d \times d$ independent pointings.  The larger possible range of supergrids generated from
random dithers yields broader distributions for the Fisher matrix elements and parameter uncertainties.  In addition, the
correlated pointings of the perfect pattern lead to interesting features when the ePSF is pixel-dominated.
Figure~\ref{fig:150} shows ${P}^{2}$, $P_{,x}^{2}$ , $PP_{,x}$ 
and $P_{,x}P_{,y}$ when the ePSF approximates a top-hat function. The function
$P_{,x}^{2}$ has two peaks separated by a pixel length; if a single pointing in
a perfect dither grid happens to have a nonzero value of $\partial(fP_{\alpha\beta})/{\partial x}$, the other pointings are guaranteed to have zero value.
This leads to highly peaked non-Gaussian distributions drawn from the partial derivatives of the ePSF.
On the other hand, the pointings of a random dither grid sample the function with no such restriction and give then smoother distributions.

Moving to a finer dither pattern by going from $2\times2$ to $3\times3$ dithers does not alter the low-uncertainty side of the distribution, but rather compresses the
high-uncertainty side to lower values.  The shifts in the mode and average of the distributions are subtle, the benefit
of going to a higher number of dither positions comes in excluding the possibility of extremely poor flux measurements.

\subsubsection{Perfect Dithering with Gaussian Pointing Accuracy}
\label{sub:GD-section}
In \S\ref{randomvperfect:sec}, it was shown that a perfect dither pattern outperforms a random pattern in certain configurations.
Practically, a perfect dither grid is impossible to obtain due to imprecisions in telescope pointing.  It is therefore useful to consider
a grid whose pointings are random realizations of an attempt to target a perfect grid.
In the following analysis the pointing
error is taken to be Gaussian-distributed with standard deviation $s_{G}$. We explore how $s_{G}$ degrades perfect dithering up to the point
where pointing error is comparable with the pixel size and the dither grid effectively becomes random.
Despite the pointing error, the realized pointings are assumed to be well determined from astrometric calibration.

There are at least two ways to model telescope pointing errors.
The first is to have independent errors for each pointing.  The second has each pointing error applied relative to the previously realized pointing, as would occur
if the grid is realized by applying a series of relative offsets.
We have explored both cases and find that they give similar results; the following analysis is based on independent pointing errors.

\begin{figure}[!t]
\center
\includegraphics[scale=0.40]{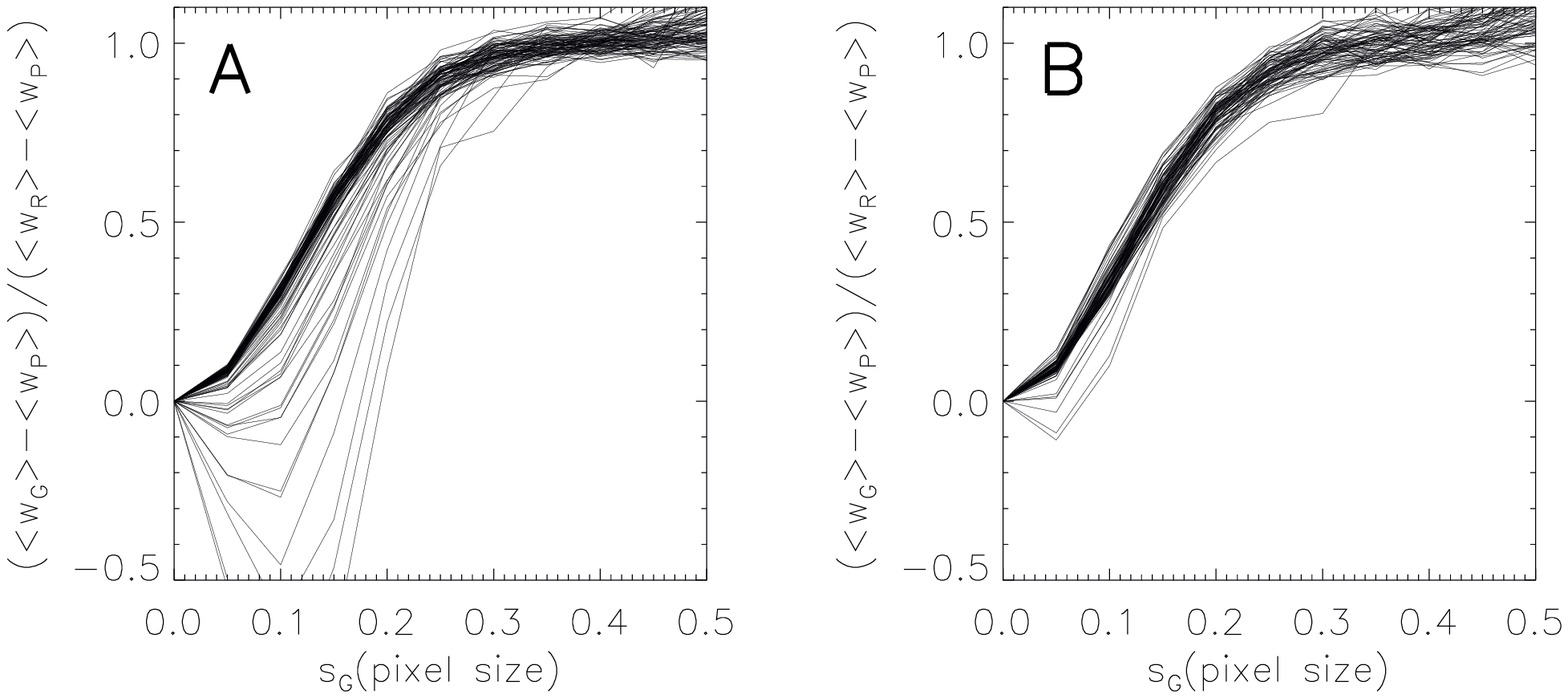}
\includegraphics[scale=0.40]{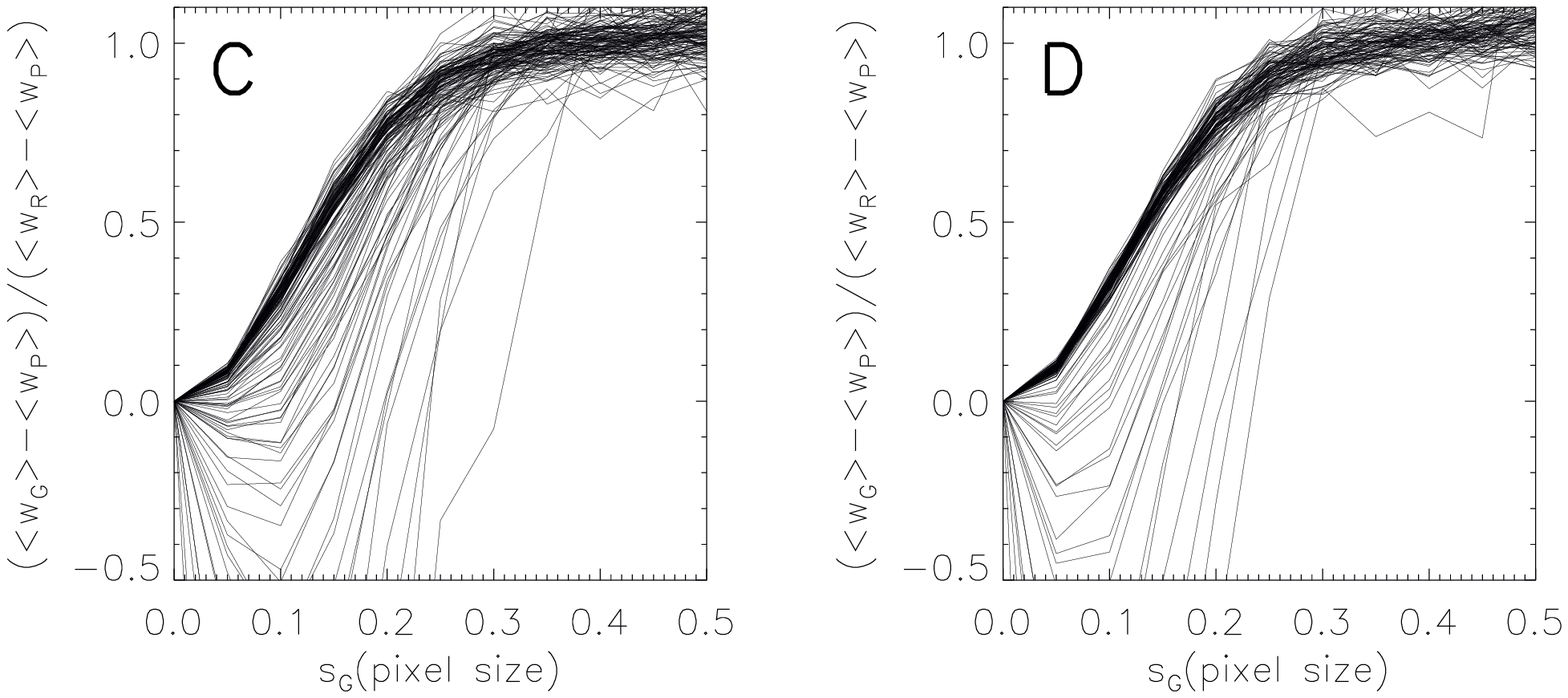}
\includegraphics[scale=0.40]{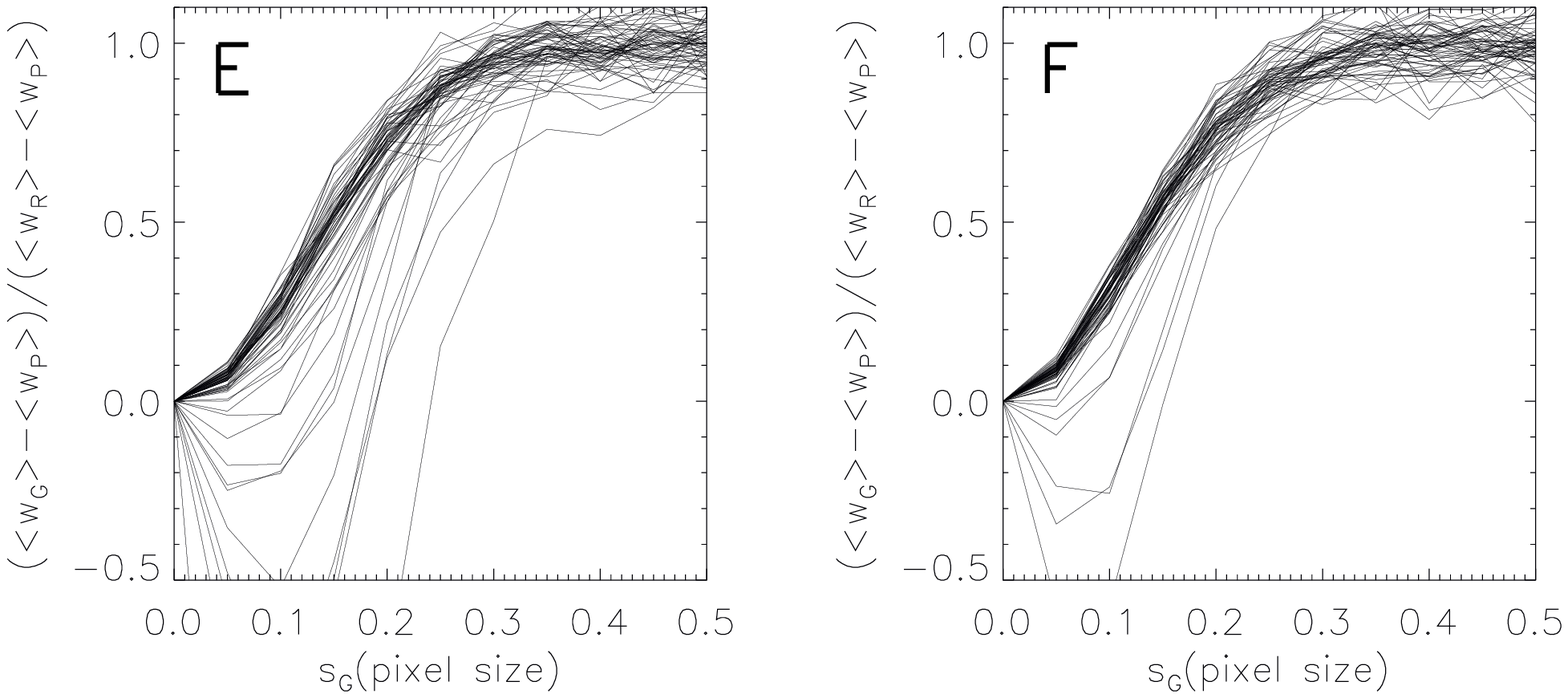}
\caption{{\footnotesize Comparison of the average parameter determinations between a perfect dither grid and the same perfect grid with Gaussian pointing errors.
Each column shows (\emph{from the top}) the ratio $(\left\langle w_{G}\right\rangle -\left\langle w_{P}\right\rangle )/(\left\langle w_{R}\right\rangle -\left\langle w_{P}\right\rangle )$ 
as a function of the Gaussian pointing error $s_{G}$ in units of pixel size: $\sigma(f)_{sky}$ (\emph{top}), $\sigma(c)_{sky}$ (\emph{middle}), 
and $\sigma(c)_{source}$ (\emph{bottom}). \emph{Left}: $2\times 2$ dither pattern. \emph{Right}: $3\times 3$. 
The lines in all the plots are for diffusions and pixel scales where perfect dithering gives average uncertainties better than $1 \%$ 
 vs. random dithering.}}
\label{fig:35}
\end{figure}

The degradation in the average measurement uncertainties from pointing errors are shown in
Figure \ref{fig:35}. The figure
 shows  the ratio $(\left\langle w_{G}\right\rangle -\left\langle w_{P}\right\rangle )/(\left\langle w_{R}\right\rangle -\left\langle w_{P}\right\rangle )$
as a function of the Gaussian pointing error $s_{G}$ at the  noise limits we consider 
for a $2\times 2$ dither pattern (left column) and a  $3\times 3$ dither pattern (right column). It shows those pixel scales and diffusions 
 where perfect dithering is better than random: $\left\langle w_{P}\right\rangle/\left\langle w_{R}\right\rangle-1<-0.01$.

For some configurations with $\left\langle w_{P}\right\rangle /\left\langle w_{R}\right\rangle -1<-0.01$ there are regions of
pointing error where the curves become negative;
slight Gaussian dispersion around  the perfect dither grid reduces average parameter uncertainties relative to both perfect and random dithering.

For flux and both position uncertainties, a pointing precision better than $\sim 0.15a$ is required to reduce $\left\langle w\right\rangle$
to get half of the advantage of a perfect dither pattern for both $2\times2$ and  $3\times3$ dithering.
 
\section{COSMIC RAYS}
\label{cosmic:sec}
The data from destructive reads from a pixel that is hit by a cosmic ray in an exposure is rendered useless, resulting in a loss of information
and reducing the accuracy of PSF photometry.
This occurs for two reasons: the integration time of the
contaminated exposure is lost and the loss of a pixel diminishes the spatial sampling of the source. 

Dithering reduces sensitivity to cosmic rays, since the probability a pixel has a cosmic-ray hit in a readout falls
with the shortened exposure time.  In the low cosmic-flux limit, with low odds that an individual pixel is hit by multiple cosmic rays,
the hit rate reduces by factor of $d\times d$ for each point in the supergrid.
Dithering also increases the number of pixels that sample the source, so that subpixel structure can be resolved
from the other pointings.

We introduce a probability
$p_{CR}$ that a cosmic ray hits
a pixel using a $1\times1$ pattern in the total exposure time
$t_{tot}$.  Incorporating cosmic-ray hits requires removing random elements in the sum over pixels
in the calculation of the Fisher elements in equation~\ref{eq:10}.
(For simplicity, we assume only one pixel is affected by a cosmic ray.)
If exposure times were the only effect, $\left\langle F\right\rangle$ would scale by the probability that the pixel is not hit
$p_{d}=1-p_{CR}/(d\times d)$
so
$\left\langle w_{CR}\right\rangle \approx\left\langle w\right\rangle/\sqrt{p_{d}} $.
Deviations from this scaling are due to the loss of spatial information in fitting the data to the ePSF.

\begin{figure}[!t]
\center
\includegraphics[scale=0.40]{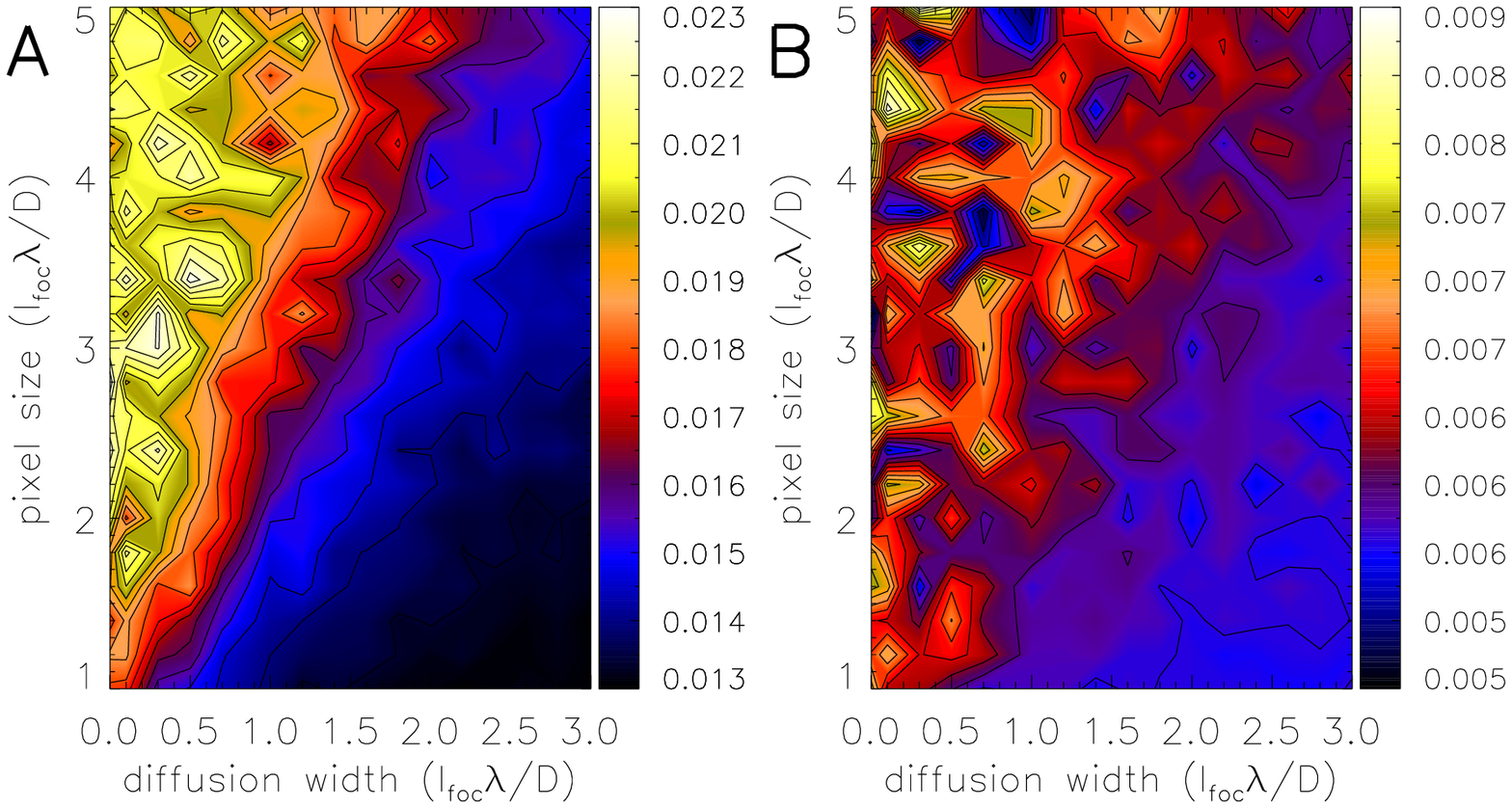}

\includegraphics[scale=0.40]{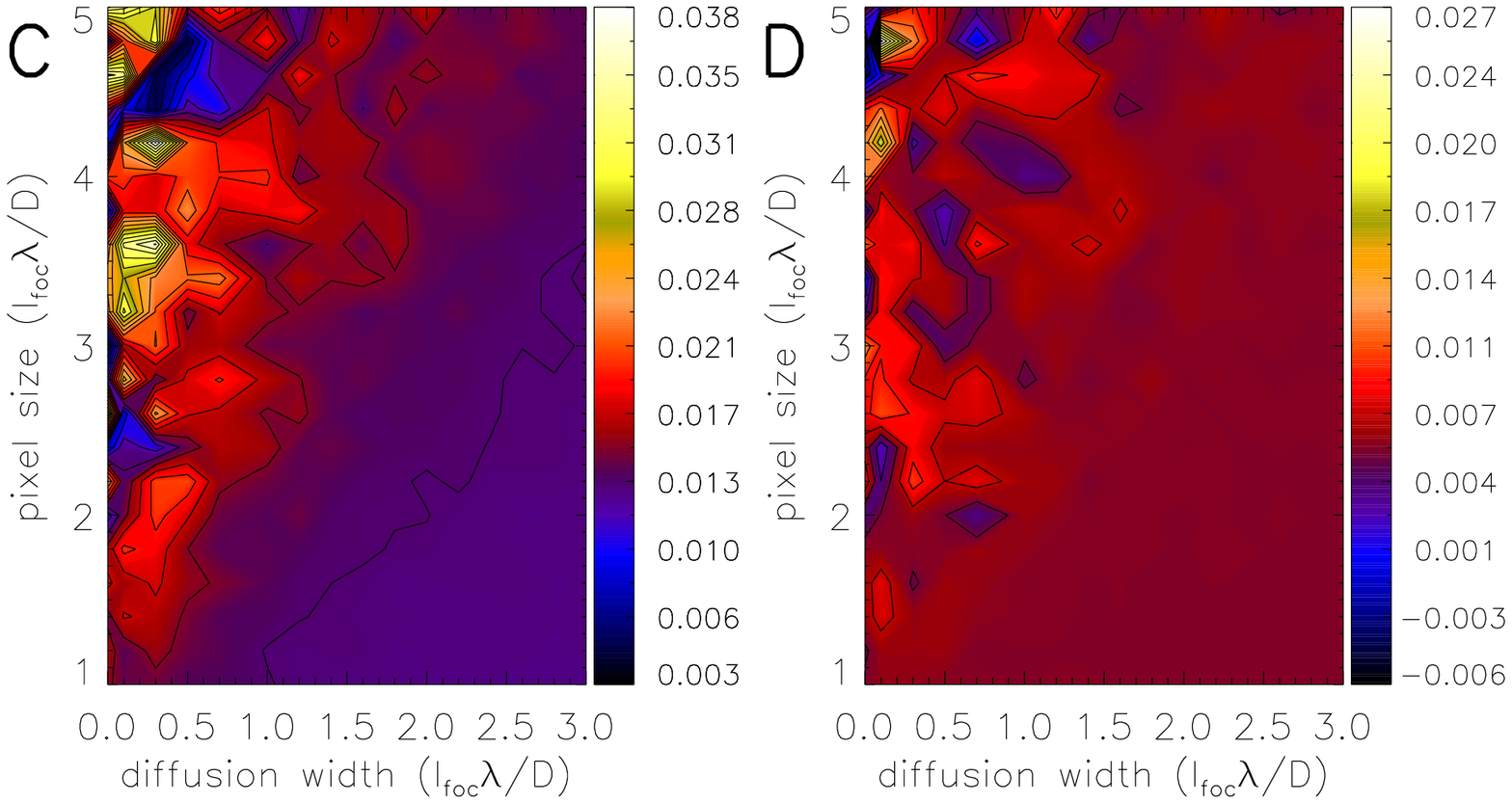}

\includegraphics[scale=0.40]{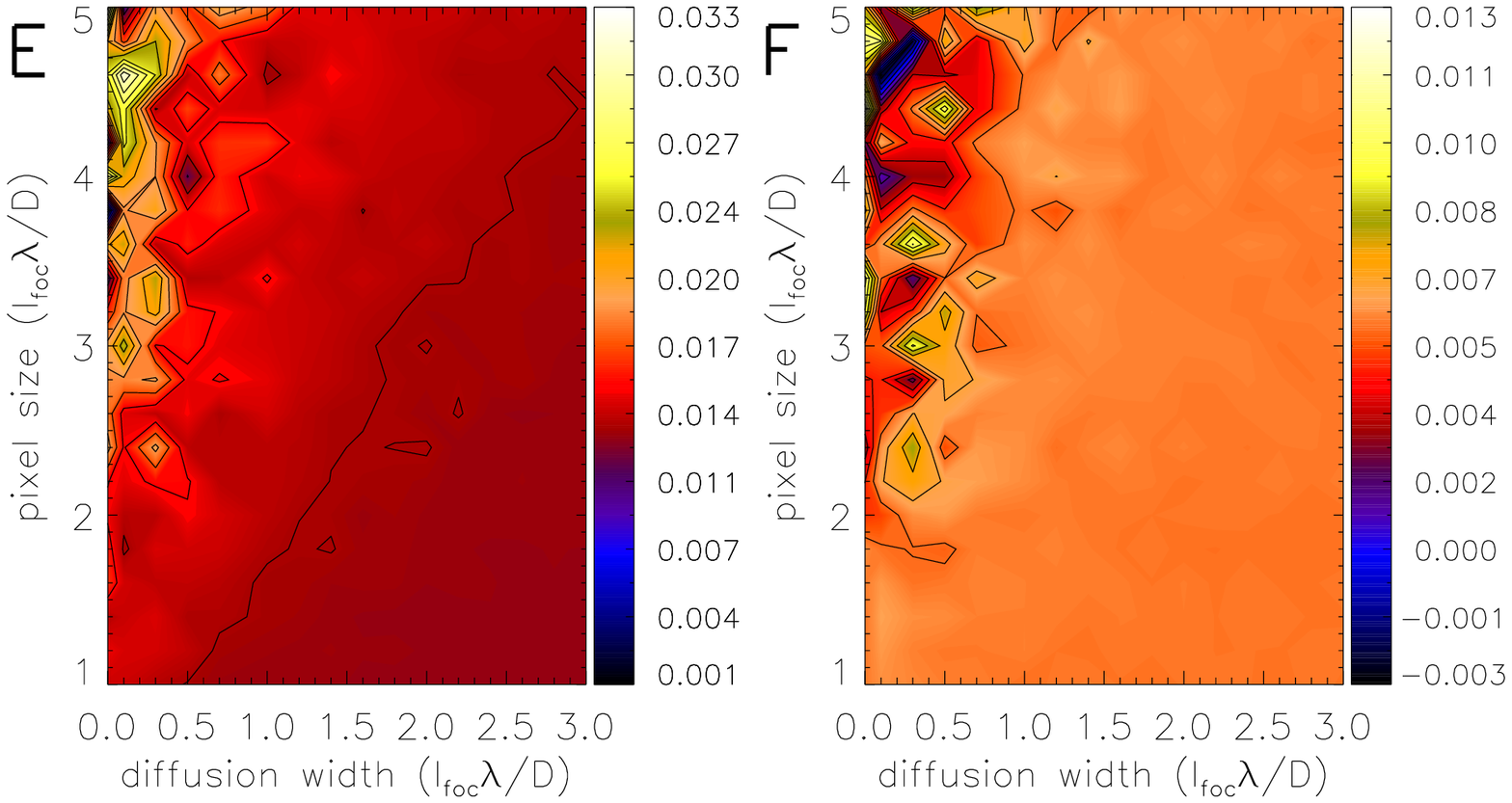}

\caption{{\footnotesize Comparison between the average parameter uncertainties with and without cosmic rays.  The cosmic-ray rate corresponds to 
a 10\% chance that a pixel is hit within the total exposure time.   The three rows correspond to the
flux uncertainties in the sky-noise limit and position uncertainties
in the sky-
and source-noise limits:  $\left\langle w_{R,CR}\right\rangle /\left\langle w_{R}\right\rangle -1$
for $\sigma(f)_{sky}$ (\emph{top}),  $\sigma(c)_{sky}$ (\emph{middle}),
and $\sigma(c)_{source}$ (\emph{bottom}). \emph{Left}: $2\times2$ dither and the \emph{Right}: $3\times3$ dither.}}
\label{fig:90}
\end{figure}

Figure \ref{fig:90}
 shows a plot of $\left\langle w_{R,CR}\right\rangle /\left\langle w_{R}\right\rangle -1$ with $p_{CR}=0.1$ for $2\times2$ and $3\times3$ dithering.
The first-order estimates $1/\sqrt{p_{d2}}-1= 0.013$ and $1/\sqrt{p_{d3}}-1=0.006$ describe 
the results when the pixels critically sample the PSF. 
In this case, the conclusions based on figures such as Figures \ref{fig:20} and \ref{fig:35}
 are still valid, but $stdev{\left(w\right)}$ and $\left\langle w\right\rangle$ increase.   When undersampled, however, there is a clear degradation
 in the average measurements beyond that accounted for by the loss in total exposure time, particularly in the $2\times2$ dither case.  When
 the ePSF is pixel-dominated, only a small number of pixels inform the fit (in the extreme case four pixels for a $2\times2$ dither)
so there is a limited draw for the distribution of realized uncertainties.
 
\section{CONCLUSIONS}
\label{conclusions:sec}
We have calculated distributions of flux and position uncertainties of randomly positioned point sources for a range of ePSFs and dither strategies.

The flux uncertainties in the sky-noise limit are dependent  on the size of the ePSF.   The position uncertainties in the sky- and source-noise limits are also dependent on the PSF size, but suffer further degradation when the PSF is undersampled.  Increased dithering reduces uncertainties only in regions where the pixel dominates the ePSF,
by only a few percent for flux, but sometimes significantly for position.  Dithering has negligible effect when the ePSF is over sampled or critically sampled.

The variance of the flux-uncertainty distribution is small and constant when the pixel size is relatively small and increases as the pixel increasingly dominates the ePSF.  The increase in variance is stronger when the PSF is dominated by diffusion (a Gaussian), as compared with being diffraction-dominated (Airy disk).  Finer dithering decreases the overall level of dispersion, but maintains the same relative dependence on pixel size and diffusion width.

Perfect and random dither patterns yield small differences in the mean of the uncertainty distributions, but can have very different variances.
The full $\sigma(f)$ distributions for random and perfect dithering and for undersampled
and oversampled regimes show that random pointings give broad distributions with a single peak, whereas
the distributions for perfect dithering are much narrower, have multiple peaks, and have sharp edges.  The perfect dithering case produces less variance
without the large flux-uncertainty realizations that random dithering does, but its asymmetric distributions can have a higher mean.
Unlike with image reconstruction, a uniform dither grid is not necessarily optimal when the pixel dominates the PSF

When a perfect dither pattern gives smaller average uncertainties, the telescope  pointing accuracy must be better than $\sim 0.15$ the pixel size to maintain half of its 
advantage.

For a fixed total exposure time, a large number of dither pointings reduces the sensitivity to cosmic rays by increasing the number of independent pixel data
measurements and reducing the probability of a cosmic hit in each.  Additional improvement occurs when the ePSF is pixel-dominated, as the better sampling
of the ePSF also influences the PSF photometry.

\acknowledgments
A. G. K.  was supported by the Director, Office of Science, Office of High Energy Physics, 
of the US\ Department of Energy under contract no. DE-AC02-05CH11231 and
DE-AC02-07CH11359. J. S.  acknowledges support from the OTICON Fund and Dark Cosmology Centre, and he
thanks the Berkeley Center for Cosmological Physics and Berkeley Lab for hospitality during his stay.

\appendix
\section{Calculation of $\left\langle F\right\rangle $}
\label{appendix:sec}

The pixels of an imaging detector map onto a grid of sky positions.  The absolute
position of the grid changes with different telescope pointings, though the relative position of the grid points remains the same.  A Fisher matrix element  for the PSF photometry of a single visit (eq.~\ref{eq:10}) is the sum of a function
$\zeta$ evaluated at the pixel grid positions for all dither pointings.
The absolute pointing of a single exposure, relative to fiducial position $\mathbf{x_0}$, is defined by the offset to the first pointing $\mathbf{h}$ and the relative offset $\mathbf{g}$ set by the dither
strategy.  The Fisher element can be thus written as
\begin{equation}
F=\sum_{\alpha \in \{Pointings\}} \sum_{i \in \{Pixels\}} \zeta(\mathbf{x_i}+ \mathbf{h} + \mathbf{g_\alpha}-\mathbf{x_0}).
\end{equation}

The
mean estimate for $F$ over all observations is given by
\begin{eqnarray}
\left\langle F\right\rangle & = &\int H(\mathbf{h})d\mathbf{h} \sum_{\alpha \in \{Pointings\}} \int G_\alpha(\mathbf{g})d\mathbf{g} \sum_{i\in \{Pixels\}}\zeta(\mathbf{x_{i}}+\mathbf{h}+\mathbf{g}-\mathbf{x_0}),
\end{eqnarray}
where $H(\mathbf{h})$ is the PDF for the starting location. $G_\alpha(\mathbf{g})$ are
the PDFs of the relative dither positions;
for perfect dithering they are delta functions centered at
the dither offset, for the Gaussian pointing uncertainties they are Gaussian distributions,
for random dithers they are a  constant.

Due to the periodicity of the (infinite) supergrid and the fact that we take the starting
point to be random, $H(\mathbf{h})$ can be taken as a normalized top hat with area $a^2$.  The spacing between pixels is $a$.  Therefore, the sum over pixels and
integral can be combined so that
\begin{equation}
\left\langle F\right\rangle =\frac{1}{a^2} \sum_{\alpha \in \{Pointings\}} \int G_\alpha(\mathbf{g})d\mathbf{g} \int d\mathbf{x} \zeta(\mathbf{x}+\mathbf{g}-\mathbf{x_0}),
\end{equation}
which simplifies to
\begin{equation}
\left\langle F\right\rangle =\left(\frac{d}{a}\right)^2\int d\mathbf{y} \zeta(\mathbf{y})
\end{equation}
for a $d\times d$ dither pattern with $d^2$ pointings.  The average value of each Fisher matrix element is independent of the dither pattern $G$.

Because of symmetry in the setup, we have that $\left\langle F_{fx}\right\rangle =\left\langle F_{xf}\right\rangle =\left\langle F_{fy}\right\rangle =\left\langle F_{yf}\right\rangle $,
$\left\langle F_{yx}\right\rangle =\left\langle F_{xy}\right\rangle $
and $\left\langle F_{xx}\right\rangle =\left\langle F_{yy}\right\rangle $;
this results in four different terms (out of nine)
\begin{eqnarray}
\left\langle F_{ff}\right\rangle& =&\frac{a^{2}t^{2}}{d^{2}}\int\frac{P^{2}}{\sigma^{2}}d\mathbf{x}, \\
\left\langle F_{xx}\right\rangle &=&\frac{a^{2}f^{2}t^{2}}{d^{2}}\int\frac{P_{,x}^{2}}{\sigma^{2}}d\mathbf{x}, \\
\left\langle F_{xf}\right\rangle &=&\frac{fa^{2}t^{2}}{d^{2}}\int\frac{PP_{,x}}{\sigma^{2}}d\mathbf{x} =0,\\
\left\langle F_{xy}\right\rangle &=&\frac{a^{2}f^{2}t^{2}}{d^{2}}\int\frac{P_{,x}P_{,y}}{\sigma^{2}}d\mathbf{x} =0,
\end{eqnarray}
where $P_{,i}$ indicates the derivative of $P$ with respect
to variable $i$. The last two terms are zero because we integrate
over the product of an odd and even function.  Therefore,
$\left\langle F\right\rangle $ is diagonal with only two different nonzero terms, 
which gives a simple expression for the inverse  $\left\langle F\right\rangle ^{-1}=1/\left\langle F\right\rangle $.

\end{document}